\documentclass[review,11pt,3p]{elsarticle}

\usepackage{amsmath,amsfonts,amssymb}
\usepackage{amsthm}
\usepackage{bm}
\usepackage{graphicx,epstopdf}
\usepackage[position=top,labelfont=normalfont,textfont=normalfont,singlelinecheck=off,justification=raggedright]{subfig}
\usepackage{floatrow}
\usepackage[dvipsnames]{xcolor}
\usepackage{setspace}
\usepackage{enumerate,etaremune}
\usepackage{booktabs}
\usepackage{tabularx,multirow}
\usepackage{framed}
\usepackage{hhline}
\usepackage{url}
\usepackage{import}
\usepackage[unicode,bookmarks=false]{hyperref}
\hypersetup{
  colorlinks,
  citecolor=Blue,linkcolor=Blue,urlcolor=Blue
}

\usepackage{lineno}

\usepackage[textsize=footnotesize,linecolor=red,backgroundcolor=red!25,bordercolor=red]
  {todonotes}

\usepackage{tikz}
\usetikzlibrary{shapes.geometric}
\usetikzlibrary{shapes,arrows}
\usetikzlibrary{plotmarks}

\usetikzlibrary{calc}

\usepackage{algorithm}
\usepackage{algorithmicx,algpseudocode}
\usepackage{cases}

\newcommand{\eg}{{\it e.g.,}}
\newcommand{\ie}{{\it i.e.,}}

\newcommand{\tensor}[1]{\bm{#1}}
\newcommand{\stress}{\sigma}
\newcommand{\strain}{\epsilon}
\newcommand{\tstress}{\tensor{\stress}}
\newcommand{\tstrain}{\tensor{\strain}}

\newcommand{\dd}{\mathrm{d}}
\newcommand{\pd}{\partial}

\newcommand{\el}{\mathrm{e}}

\newcommand{\rn}[1]{\uppercase\expandafter{\romannumeral #1\relax}}

\let\diver\relax
\DeclareMathOperator{\grad}{\nabla}
\DeclareMathOperator{\symgrad}{\nabla^{s}}

\DeclareMathOperator{\diver}{\grad\cdot}

\DeclareMathOperator{\dyadic}{\otimes}
\newsavebox{\dotbox}

\theoremstyle{remark}

\newcolumntype{L}[1]{>{\raggedright\let\newline\\arraybackslash\hspace{0pt}}m{#1}}
\newcolumntype{C}[1]{>{\centering\let\newline\\arraybackslash\hspace{0pt}}m{#1}}
\newcolumntype{R}[1]{>{\raggedleft\let\newline\\arraybackslash\hspace{0pt}}m{#1}}

\allowdisplaybreaks

\biboptions{sort&compress,comma,authoryear}
\AtBeginDocument{\hypersetup{citecolor=MidnightBlue,linkcolor=MidnightBlue,urlcolor=MidnightBlue}}

\usepackage{etoolbox}
\makeatletter
\patchcmd{\ps@pprintTitle}
{Preprint submitted to}
{~}
{}{}
\makeatother

\begin{document}

\begin{frontmatter}

\title{Crack opening calculation in phase-field modeling of fluid-filled fracture:\\ A robust and efficient strain-based method}

\author[LLNL]{Fan Fei}
\author[KAIST]{Jinhyun Choo\corref{corr}}
\cortext[corr]{Corresponding Author}
\ead{jinhyun.choo@kaist.ac.kr}

\address[LLNL]{Atmospheric, Earth, and Energy Division, Lawrence Livermore National Laboratory, Livermore, CA, USA}
\address[KAIST]{Department of Civil and Environmental Engineering, KAIST, Daejeon, South Korea}
\journal{~}

\begin{abstract}
The phase-field method has become popular for the numerical modeling of fluid-filled fractures, thanks to its ability to represent complex fracture geometry without algorithms. However, the algorithm-free representation of fracture geometry poses a significant challenge in calculating the crack opening (aperture) of phase-field fracture, which governs the fracture permeability and hence the overall hydromechanical behavior. Although several approaches have been devised to compute the crack opening of phase-field fracture, they require a sophisticated algorithm for post-processing the phase-field values or an additional parameter sensitive to the element size and alignment. Here, we develop a novel method for calculating the crack opening of fluid-filled phase-field fracture, which enables one to obtain the crack opening without additional algorithms or parameters. We transform the displacement-jump-based kinematics of a fracture into a continuous strain-based version, insert it into a force balance equation on the fracture, and apply the phase-field approximation. Through this procedure, we obtain a simple equation for the crack opening which can be calculated with quantities at individual material points. We verify the proposed method with analytical and numerical solutions obtained based on discrete representations of fractures, demonstrating its capability to calculate the crack opening regardless of the element size or alignment.
\end{abstract}

\begin{keyword}
Phase-field method \sep
Crack opening \sep
Fluid-filled fracture \sep
Hydraulic fracture \sep
Poromechanics \sep
Coupled problems
\end{keyword}
 
\end{frontmatter}


\section{Introduction}
\label{sec:intro}

The hydromechanical behavior of fluid-filled fractures is pivotal to the performance of many subsurface technologies for energy and the environment.
Examples include unconventional resource recovery, enhanced geothermal systems, radioactive waste disposal, and geologic carbon storage. 
Numerical modeling plays a critical role in addressing hydromechanical processes in fluid-filled fractures in these applications. 

In recent years, the phase-field method has become a popular approach for numerical modeling of fluid-filled fractures (\eg~\cite{bourdin2012variational,mikelic2015phase,choo2018cracking,lee2016pressure,ha2018liquid, santillan2018phase,chukwudozie2019variational,fei2023phaseb}). 
Unlike discrete approaches that treat fractures as lower-dimensional entities (\eg~\cite{khoei2014mesh,settgast2017fully,cusini2021simulation}), the phase-field method approximates fracture geometry diffusely with a continuous field variable, namely, the phase field.  
In this way, the phase-field method can represent the fracture geometry without any algorithms, allowing one to model geometrically complex fractures far more easily than the discrete approaches. 
This feature is particularly appealing for modeling evolving fractures.

When it comes to modeling fluid flow in fracture, however, the phase-field method can be more complex than the discrete approaches.
This is because, for modeling fluid flow in fracture, one has to calculate the crack opening (aperture) to estimate the fracture permeability using the cubic law or similar equations.
In the discrete approaches, the crack opening can be calculated explicitly as the jump in the normal displacement across the fracture.
In the phase-field method, however, the crack opening cannot be calculated straightforwardly because the phase-field method represents the fracture implicitly as a diffuse interface.
Therefore, despite its many advantages for other purposes, the algorithm-free representation of fracture geometry in the phase-field method gives rise to a significant challenge for modeling fluid flow in fracture.

In the literature, several approaches have been devised to calculate the crack opening of a fluid-filled fracture described by the phase-field method. 
These approaches may be classified into two categories: 
(i) displacement-jump-based approaches, whereby the displacement jump across the phase-field fracture is computed via post-processing of the phase-field values, 
and (ii) strain-based approaches, whereby the crack opening is calculated from the strain tensor inside the phase-field fracture (\ie~the smeared fracture zone).

The most well-known displacement-jump-based approach is the line integral method, which was used in \cite{bourdin2012variational} and later adopted and analyzed in several other studies~\citep{santillan2018phase,chukwudozie2019variational,yoshioka2020crack}.
Recently, \cite{chen2023computation} have also introduced a revised version of the line integral method where an approximated Dirac delta function is used to compute the crack opening of cohesive fractures. 
Meanwhile, \cite{lee2017iterative} have proposed a different method for calculating the crack opening of a phase-field fracture, which utilizes the level-set method to distinguish between the fracture and matrix.
Other researchers have calculated the crack opening by constructing discrete representations of phase-field fractures~\citep{sun2020hybrid,costa2022multi,yang2021explicit}. 
While these displacement-jump-based approaches are sufficiently accurate, they require significant effort for implementation and computation as they involve sophisticated algorithms to post-process the phase-field values. 

Alternatively, some phase-field models calculated the crack opening based on the strain tensor inside the phase-field fracture (\eg~\cite{miehe2015minimization,wilson2016phase,you2023poroelastic}).
Such strain-based methods are far more efficient and easier to implement than displacement-jump-based methods.
To convert the strain tensor into the crack opening, the existing strain-based approaches commonly multiply a scalar strain to the characteristic length (size) of the fractured element. 
As shown in previous studies~\citep{wilson2016phase,you2023poroelastic}, this calculation is valid when the following two conditions are met: (i) the crack opening is sufficiently small compared with the element size, and (ii) the phase-field fracture is aligned with the element direction.
However, if either of these two conditions is unsatisfied---often the case for complex fractures---the existing strain-based approaches become erroneous. 
(See~\cite{wilson2016phase} for some numerical evidence.)
Therefore, despite their simplicity, the existing strain-based approaches may not be suitable for complex fracture problems for which the phase-field method can be more attractive than discrete methods.
In this regard, it would be prudent to conclude that no existing method allows one to calculate the crack opening in phase-field modeling of fluid-filled fractures with robustness and efficiency.

In this paper, we develop a novel strain-based method that can accurately and efficiently calculate the crack opening in phase-field modeling of fluid-filled fracture.
To this end, we first formulate a strain-based kinematic description of an open fracture filled with fluid, building on the way how recent phase-field models of geologic fracture~\citep{fei2022phase,fei2023phase} transform displacement-jump-based kinematics into a continuous strain-based version. 
Inserting this strain-based kinematic description into a force balance equation on the fracture and applying the phase-field approximation, we obtain an equation for the crack opening in the phase-field setting.
The equation can be calculated at individual material points without a characteristic length parameter.
We then verify the proposed method with analytical and numerical results obtained based on discrete representations of fractures.
To our knowledge, this is the first method that enables one to accurately calculate the crack opening in phase-field modeling of fluid-filled fracture, without a sophisticated algorithm for post-processing the phase-ﬁeld values or an additional parameter sensitive to the element size and alignment.

The paper proceeds as follows.
In Section~\ref{sec:formulation}, we recapitulate a standard phase-field model for fluid-filled fracture in porous media. 
In Section~\ref{sec:calculation}, we derive a new formulation for calculating the crack opening of a phase-field fracture filled with fluid, which is the main contribution of this work. 
In Section~\ref{sec:implementation}, we describe how to implement the proposed method during the numerical solution of the phase-field model.
In Section~\ref{sec:verification}, we verify the proposed method with reference solutions obtained by analytical and other discrete numerical methods. 
In Section~\ref{sec:closure}, we conclude the paper.

\section{Phase-field formulation for fluid-filled fracture}
\label{sec:formulation}

This section presents a standard phase-field formulation for modeling fluid-filled fracture in porous media. 

\subsection{Phase-field approximation}

Consider a fluid-saturated porous domain $\Omega \in \mathbb{R}^{n_\text{dim}}$ where $n_\text{dim}$ denotes the spatial dimension.
Its exterior boundary, $\pd \Omega$, is decomposed into a Dirichlet part and a Neumann part, namely, $\pd_{u} \Omega$ and $\pd_{t} \Omega$ for the solid deformation problem, and $\pd_{p} \Omega$ and $\pd_{q} \Omega$ for the fluid flow problem, such that $\pd_{u} \Omega \cup \pd_{t} \Omega = \pd_{p} \Omega \cup \pd_{q} \Omega = \pd \Omega$ and $\pd_{u} \Omega \cap \pd_{t} \Omega = \pd_{p} \Omega \cap \pd_{q} \Omega = \emptyset$.
The domain contains a set of lower dimensional fractures, denoted by $\Gamma$, which are assumed to be filled with a nearly incompressible fluid such as water.
Also, the time domain of interest is denoted by $\mathbb{T}:= [0, t_{\max}]$. 

\begin{figure}
    \centering
    \includegraphics[width=\textwidth]{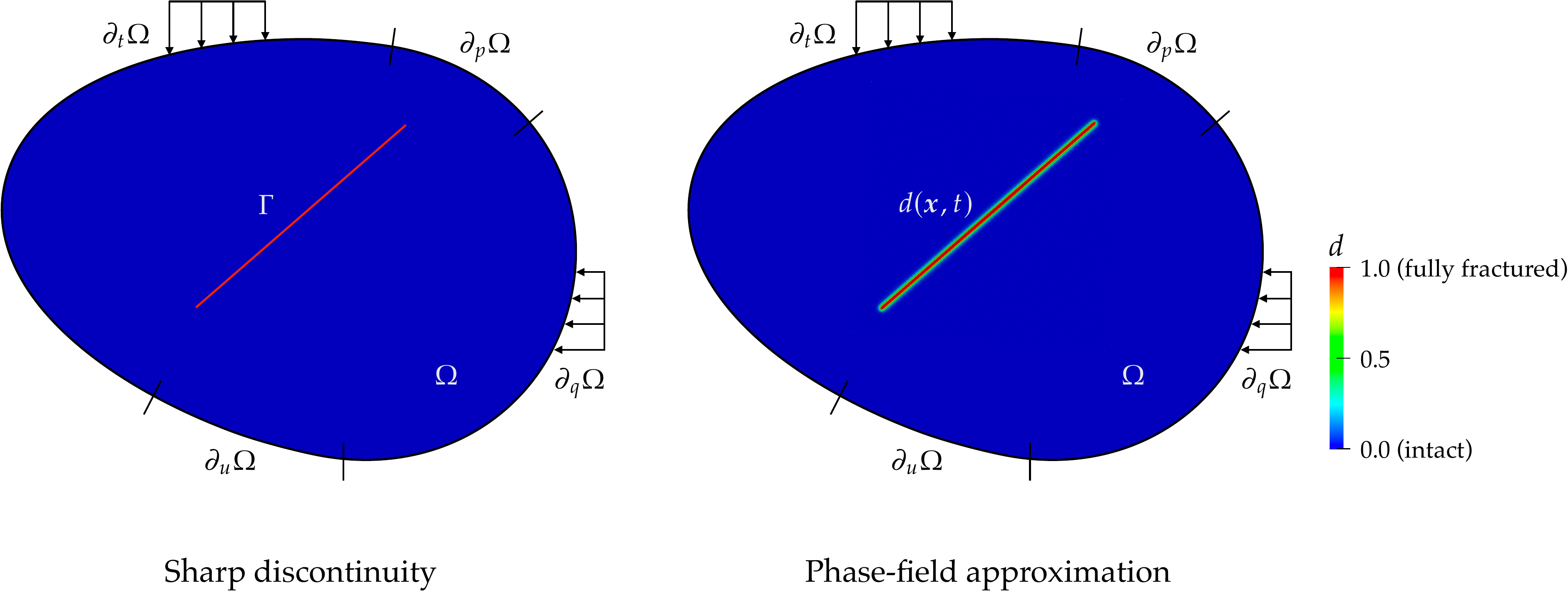}
    \caption{Phase-field approximation of fracture. The sharp discontinuity $\Gamma$ is diffusely approximated by the phase-field variable, $d\in[0,1]$.}
    \label{fig:pf-approximate}
\end{figure}

Figure~\ref{fig:pf-approximate} illustrates how the phase-field variable, $d \in [0,1]$, diffusely approximates the sharp discontinuity $\Gamma$.
The region with $d = 0$ corresponds to an intact region, whereas the region with $d = 1$ corresponds to a fully fractured region.
So the phase-field variable may also be interpreted as a damage variable, although its evolution is usually described by fracture mechanics theory.
Mathematically, the phase-field approximation is accomplished by introducing a crack density functional, $\Gamma_{d}(d, \grad d)$, satisfying 
\begin{align}
   \int_{\Gamma} \: \dd A = \int_\Omega \delta_{\Gamma} (\tensor{x}) \: \dd V \approx \int_{\Omega} \Gamma_{d}(d, \grad d) \: \dd V, \label{eq:gamma-approximate}
\end{align}
where $\delta_\Gamma(\tensor{x})$ denotes the Dirac delta function, defined as
\begin{align}
    \delta_{\Gamma}(\tensor{x}) = 
    \left \{
    \begin{array}{ll}
        1 & \text{on} \,\, \Gamma,   \vspace{0.5em} \\
        0 & \text{in} \,\, \Omega \backslash \Gamma.
    \end{array}
    \right . 
\end{align}
In the literature, a few different forms of $\Gamma_{d}(d, \grad d)$ have been used.
In this work, we use the most widely used form of $\Gamma_{d}(d, \grad d)$, given by
\begin{align}
    \Gamma_{d}(d,\grad d) := \dfrac{1}{2L}\left(d^{2} + L^2 \grad d \cdot \grad d \right),  
    \label{eq:gamma_d}
\end{align}
where $L$ is the phase-field regularization length that controls the width of the diffuse zone. 
Note that the crack opening calculation method proposed in this work is independent of the choice of $\Gamma_{d}(d, \grad d)$, and hence it can be applied to other forms of $\Gamma_{d}(d, \grad d)$ such as those employed cohesive phase-field fractures (\eg~\cite{geelen2019phase,wu2017unified,fei2020phaseb}). 
Here, we have chosen the form in Eq.~\eqref{eq:gamma_d} because it is most common in the existing phase-field models of fracture. 

\subsection{Governing equations}
A phase-field formulation for fluid-filled fracture is furnished by three governing equations where the primary variables are the displacement field, $\tensor{u}$, the phase field, $d$, and the fluid pressure field, $p$.
Because this work focuses on the calculation of the crack opening of a phase-field fracture, here we adopt the governing equations derived in the literature, omitting their derivation for brevity. 
Let us assume that the fluid pressure field is continuous across the fracture--matrix interface. 
In this case, the momentum balance equation is given by~\citep{santillan2018phase}
\begin{equation}
    \diver \left[\tstress' - b g(d) p \tensor{1} \right] + p\grad g(d) + \rho \tensor{g} = \tensor{0} \quad \text{in} \;\; \Omega \times \mathbb{T}. 
    \label{eq:momentum-balance}
\end{equation}
Here, $\tstress'$ is the effective stress tensor, $b$ is the Biot coefficient, $g(d)$ is a degradation function that accounts for stiffness reduction in the fractured zone, $\rho$ is the mass density of the porous medium, and $\tensor{g}$ is the gravitational acceleration vector. 
Assuming that the solid matrix is isotropic and linear elastic, the effective stress tensor can be constitutively related to the (infinitesimal) strain tensor, $\tstrain :=\symgrad \tensor{u}$, as 
\begin{equation}
    \tstress' = g(d) \mathbb{C}^{\el}:\tstrain,
\end{equation} 
where
\begin{equation}
    \mathbb{C}^{\el} = \lambda \mathbb{I} + 2\mu \tensor{1} \dyadic \tensor{1}.
\end{equation}
Here, $\lambda$ and $\mu$ are the Lam\'{e} parameters, which are related to other elasticity parameters including Young's modulus $E$ and Poisson's ratio $\nu$, and $\mathbb{I}$ and $\tensor{1}$ are the fourth-order and second-order identity tensors, respectively. 
As for the degradation function, we choose the most popular quadratic form, namely, $g(d) = (1 - d)^2$.
The boundary conditions for the momentum balance equation are given by 
\begin{align}
    \tensor{u} = \hat{\tensor{u}} \quad &\text{on} \;\; \pd_{u} \Omega \times \mathbb{T},\\
    [\tstress' - b g(d) p \tensor{1}] \cdot\bar{\tensor{n}}  = \hat{\tensor{t}} \quad &\text{on} \;\; \pd_{t} \Omega \times \mathbb{T}
\end{align}
where $\hat{\tensor{u}}$ and $\hat{\tensor{t}}$ are the boundary displacement and traction vectors, respectively, and $\bar{\tensor{n}}$ is the normal vector to the exterior boundary. 

The phase-field governing equation is given by
\begin{equation}
    -g'(d) \left[ \mathcal{H}^{+}(\tstrain,t) + (1 - b) p \diver \tensor{u} + \tensor{u} \cdot \grad p \right] = \dfrac{\mathcal{G}_{c}}{L} \left(d - L^2 \diver \grad d \right)  \quad \text{in} \;\; \Omega \times \mathbb{T},
    \label{eq:damage-eq-H+}
\end{equation}
where $\mathcal{H}^{+}(\tstrain,t)$ is the crack driving force related to the tensile strain energy, and $\mathcal{G}_{c}$ is the critical fracture energy. 
As such, the phase-field governing equation describes the energy balance among the released tensile strain energy, the work done by the fluid pressure in the matrix and fracture, and the dissipation due to fracture propagation.
It is noted that $\mathcal{H}^{+}(\tstrain,t)$ is set as the maximum tensile strain energy in history to enforce crack irreversibility~\citep{miehe2010phase}.
At the domain boundary, the phase-field variable is subjected to the homogeneous Neumann boundary condition, namely,
\begin{equation}
    \grad d \cdot \bar{\tensor{n}} = 0 \:\:\text{on} \;\; \pd \Omega \times \mathbb{T}.
\end{equation}

The governing equation for fluid flow is provided by the balance of fluid mass.
As common in geomechanics, let us assume that the solid grain is incompressible.
Under this assumption, the mass balance equation for a fractured porous medium can be written as~\citep{you2023poroelastic}
\begin{equation}
    \dfrac{\bar{\phi}}{K_{f}}\dot{p} - \bar{b} \diver \dot{\tensor{u}} +  \diver \tensor{q} = Q \quad \text{in} \;\; \Omega \times \mathbb{T}. \label{eq:flow-pf-fracture-general}
\end{equation}
where dot denotes the material time derivative, $K_{f}$ denotes the bulk modulus of the pore fluid, $\bar{b}$ and $\bar{\phi}$ denote the effective Biot coefficient and porosity, respectively, of the fractured porous medium, $\tensor{q}$ denotes the superficial fluid flux vector, and $Q$ denotes the source/sink term. 
According to~\cite{you2023poroelastic}, $\bar{b}$ and $\bar{\phi}$ can be related to the phase-field variable as
\begin{align}
    \bar{b}(d) &= 1 - g(d) ( 1- b), \\
    \bar{\phi}(d) &= 1 - g(d) ( 1- \phi_{m}),
\end{align}
where $\phi_{m}$ is the porosity of the intact rock matrix.  
Also, assuming Darcy's law is valid for fluid flow in both the matrix and fracture regions, the superficial fluid flux can be evaluated as 
\begin{equation}
    \tensor{q} = -\dfrac{\tensor{k}}{\mu_{f}} \left( \grad p - \rho \tensor{g} \right). 
\end{equation}
Here, $\mu_{f}$ is the fluid viscosity, and $\tensor{k}$ is the homogenized permeability tensor of the fractured porous medium, given by
\begin{equation}
    \tensor{k} = k_{m} \tensor{1} + m(d)\tensor{k}_{f}, 
\end{equation} 
where $k_{m}$ is the matrix permeability, $\tensor{k}_{f}$ is the fracture permeability tensor, and $m(d)$ is a weighting function satisfying $m(0) = 0$ and $m(1) = 1$. 
For simplicity, we use $m(d) = 1 - g(d)$ in this work. 
As for the fracture permeability tensor, it can be written based on the cubic law as
\begin{equation}
    \tensor{k}_{f} = \dfrac{\omega^{2}}{12}(\tensor{1} - \tensor{n} \dyadic \tensor{n}),
\end{equation} 
where $\omega$ is the crack opening (aperture).
It can be seen from the above equation that the calculation of the crack opening is central to the evaluation of the fracture permeability tensor.
Lastly, the Dirichlet and Neumann boundary conditions for the fluid flow problem are given by
\begin{align}
    p = \hat{p} \quad &\text{on} \;\; \pd_{p} \Omega \times \mathbb{T}, \\ 
    \tensor{q} \cdot \tensor{n} = \hat{q} \quad &\text{on} \;\; \pd_{q} \Omega \times \mathbb{T},
\end{align}
where $\hat{p}$ and $\hat{q}$ are the prescribed pressure and flux, respectively, on the boundary.

\section{Crack opening calculation}
\label{sec:calculation}

In this section, we develop a new equation for the crack opening of a fluid-filled fracture modeled by the phase-field method.

\subsection{Strain-based kinematics of fracture}
Following the strong discontinuity approach~\citep{simo1993analysis,borja2000finite}, we decompose the displacement field into the continuous part, $\bar{\tensor{u}}$, and the discontinuous part (displacement jump), $[\![ \tensor{u}]\!] $, as
\begin{align}
    \tensor{u} := \bar{\tensor{u}} + H_{\Gamma}(\tensor{x})[\![ \tensor{u}]\!] .  
    \label{eq:disp-decompose}
\end{align}
Here, $H_{\Gamma}(\tensor{x})$ denotes the Heaviside function, defined as
\begin{align}
    H_{\Gamma}(\tensor{x}) = 
    \left \{
    \begin{array}{ll}
        1 & \text{if} \,\, \tensor{x} \in \Omega_{+},   \vspace{0.5em} \\
        0 & \text{if} \,\, \tensor{x} \in \Omega_{-},
    \end{array}
    \right . 
\end{align}
where $\Omega_{+}$ and $\Omega_{-}$ refer to the subdomains on the positive and negative sides, respectively, of the fracture.
When the fracture is filled with fluid, it can be assumed that only the normal component of the displacement field is discontinuous across the fracture.
In this case, the displacement jump can be expressed as
\begin{align}
	[\![ \tensor{u} ]\!] =  \omega \tensor{n}. \label{eq:opening}
\end{align} 
Inserting Eq.~\eqref{eq:opening} into Eq.~\eqref{eq:disp-decompose} gives 
\begin{align}
	\tensor{u} := \bar{\tensor{u}} + H_{\Gamma}(\tensor{x})\omega \tensor{n} .  \label{eq:disp-decompose-full}
\end{align}
Taking the symmetric gradient of Eq.~\eqref{eq:disp-decompose-full}, we obtain the strain tensor in the fracture as
\begin{align}
	\tstrain := \grad^\mathrm{s} \tensor{u} = \bar{\tstrain} + \dfrac{1}{2} H_{\Gamma}(\tensor{x}) (\tensor{n} \dyadic \grad \omega + \grad \omega \dyadic \tensor{n})   +   \omega \delta_{\Gamma} (\tensor{x})\tensor{n} \dyadic \tensor{n}.  \label{eq:strain-decompose}
\end{align}
Here, $\bar{\tstrain} := \symgrad \bar{\tensor{u}}$ is the continuous part of the strain tensor, and $\delta_{\Gamma}(\tensor{x})$ is the Dirac delta function derived from the gradient of the Heaviside function, \ie~$\delta_{\Gamma} (\tensor{x}) := \grad H_{\Gamma}(\tensor{x})$. 

\subsection{Crack opening of fluid-filled fracture}
Based on the strain-based kinematics of fracture, we develop a method to calculate the crack opening of a phase-field fracture filled with fluid.
First, recall that the effective stress tensor must be non-singular and continuous across the fracture~\citep{regueiro2001plane}. 
So the effective stress tensor, $\tstress'$, and the continuous part of the strain tensor, $\bar{\tstrain}$, can be related as
\begin{align}
    \tstress' = \mathbb{C}^{\el}:\bar{\tstrain}.
\end{align}
Inserting Eq.~\eqref{eq:strain-decompose} into the above equation yields 
\begin{align}
    \tstress' = \mathbb{C}^{\el}:\left[\tstrain - \dfrac{1}{2} H_{\Gamma}(\tensor{x}) (\tensor{n} \dyadic \grad \omega + \grad \omega \dyadic \tensor{n})  -  \omega \delta_{\Gamma} (\tensor{x})\tensor{n} \dyadic \tensor{n}\right]. 
    \label{eq:cauthy-stress}
\end{align}
Also, the normal stress on the crack surface should be balanced with the fluid pressure inside the fracture. So we have
\begin{align}
    \left[\tstress' - bg(d) p \tensor{1} \right]\cdot\tensor{n} = -p \tensor{n} \:\:\text{on} \:\: \Gamma .  
\end{align}
Since $g(d) = 0$ in the fracture region ($d=1$), the above equation simplifies to
\begin{align}
    \tstress' \cdot\tensor{n} = -p \tensor{n} \:\:\text{on} \:\: \Gamma .  \label{eq:force-balance}
\end{align}
Combining Eq.~\eqref{eq:force-balance} with Eq.~\eqref{eq:cauthy-stress} allows us to express $p$ as
\begin{align}
    p = - \left(\lambda \tensor{1} + 2\mu \tensor{n}\dyadic \tensor{n} \right):\left[\tstrain - \dfrac{1}{2} H_{\Gamma}(\tensor{x}) (\tensor{n} \dyadic \grad \omega + \grad \omega \dyadic \tensor{n})   -   \omega \delta_{\Gamma} (\tensor{x})\tensor{n} \dyadic \tensor{n} \right] \:\:\text{on} \:\: \Gamma .
    \label{eq:pressure-omega}
\end{align}
Rearranging Eq.~\eqref{eq:pressure-omega}, we can express the crack opening, $\omega$, as
\begin{align}
    \omega = \dfrac{\left(\lambda \tensor{1} + 2\mu \tensor{n}\dyadic \tensor{n} \right):\tstrain - (\lambda + 2\mu) H_{\Gamma}(\tensor{x}) \tensor{n} \cdot \grad \omega + p}{\delta_{\Gamma}(\tensor{x}) (\lambda + 2\mu)} \:\:\text{on} \:\: \Gamma .
    \label{eq:omega-compute}
\end{align}
Let us assume that $\omega$ is constant along the crack normal direction, say, $\tensor{n} \cdot \grad \omega = 0$.
Then Eq.~\eqref{eq:omega-compute} simplifies to
\begin{align}
    \omega = \dfrac{\left(\lambda \tensor{1} + 2\mu \tensor{n}\dyadic \tensor{n} \right):\tstrain + p}{\delta_{\Gamma}(\tensor{x}) (\lambda + 2\mu)}\:\:\text{on} \:\: \Gamma . \label{eq:omega-delta}
\end{align}
This expression is yet incompatible with the phase-field formulation due to the presence of the Dirac delta function $\delta_{\Gamma}(\tensor{x})$. 
Therefore, by appealing to Eq.~\eqref{eq:gamma-approximate}, we approximate $\delta_{\Gamma}(\tensor{x})$ in Eq.~\eqref{eq:omega-delta} by the crack density functional, $\Gamma_{d} (d ,\grad d)$. 
It is noted that an approximated Dirac delta function was also used in~\cite{chen2023computation} with the line integral method.
Finally, we obtain
\begin{align}
    \omega \approx \dfrac{\left(\lambda \tensor{1} + 2\mu \tensor{n}\dyadic \tensor{n} \right):\tstrain + p}{\Gamma_{d} (d ,\grad d) (\lambda + 2\mu)} \:\: \text{in} \:\: \Omega \:\: \text{where} \:\: d > 0 .
    \label{eq:omega-final}
\end{align}    
This is the equation we propose for crack opening calculation in phase-field modeling of fluid-filled fracture.
Note that the above equation applies only inside the phase-field fracture ($d > 0$) where $\Gamma_{d}(d, \grad d)$ in the denominator is nonzero.
It is also noted that the above equation can be used for calculating the crack opening of an empty open crack by setting $p=0$.

The proposed method sets it apart from other crack-opening-calculation methods in the phase-field literature.
Unlike the existing displacement-jump-based approaches, it does not require any algorithm for line integral or fracture geometry reconstruction, which demands significant effort for implementation and computation.
Also, unlike the existing strain-based approaches, it does not introduce any characteristic length sensitive to the element size and alignment.
Without any additional parameters, the proposed method can be implemented easily during a material-point update, as described in the next section.

\section{Implementation}
\label{sec:implementation}

This section describes how to implement the proposed method in the phase-field modeling of fluid-filled fractures.
For completeness, we first describe the spatial and temporal discretization of the governing equations.
Then we describe a simple algorithm to implement the proposed method in the material update subroutine.

\subsection{Discretization}
For spatial discretization of the governing equations, we use a three-field mixed finite element method. 
The trial solution spaces for $\tensor{u}$, $d$, and $p$ are defined as
\begin{align}
    \mathcal{S}_{u} &:= \left\{ \tensor{u} \: \rvert \: \tensor{u} \in \tensor{H}^{1},\, \tensor{u} = \hat{\tensor{u}} \,\, \text{on} \,\, \pd_{u} \Omega  \right\}, \\ 
    \mathcal{S}_{d} &:= \left\{ d \: \rvert \: d \in {H}^{1} \right\}, \\
    \mathcal{S}_{p} &:= \left\{ p \: \rvert \: p \in {H}^{1},\, p = \hat{p} \,\, \text{on} \,\, \pd_{p} \Omega  \right\},
\end{align}
where $H^{1}$ denotes a Sobolev space of degree one. 
The corresponding weighting function spaces can be defined as 
\begin{align}
    \mathcal{V}_{u} &:= \left\{ \tensor{\eta} \: \rvert \: \tensor{\eta} \in \tensor{H}^{1}, \tensor{\eta} = \tensor{0} \,\, \text{on} \,\, \pd_{u} \Omega  \right\}, \\ 
    \mathcal{V}_{d} &:= \left\{ \phi \: \rvert \: \phi \in {H}^{1} \right\},  \\ 
    \mathcal{V}_{p} &:= \left\{ \psi \: \rvert \: \psi \in {H}^{1}, \psi = 0 \,\, \text{on} \,\, \pd_{p} \Omega  \right\}, 
\end{align}
For temporal discretization of the time-dependent terms (the solid velocity vector and the rate of the crack opening), we use the implicit Euler method.
Through the standard weighted residual procedure, we obtain the time-discrete variational equations in residual form as follows:
\begin{align}
    \mathcal{R}_{u}(\tensor{u},d,p) := &- \int_\Omega \left[\tstress'(\symgrad\tensor{u}) - g(d) b p \tensor{1}\right] : \symgrad \tensor{\eta} \: \dd V + \int_\Omega p \grad g(d) \cdot \tensor{\eta} \: \dd V  
    + \int_\Omega \rho \tensor{g} \cdot \tensor{\eta} \: \dd V \nonumber \\ 
    &+ \int_{\pd_{t} \Omega} \hat{\tensor{t}} \cdot \tensor{\eta} \: \dd A = 0, 
    \label{eq:momentum-balance-residual} \\ 
    \mathcal{R}_{d}(\tensor{u},d,p)  := &\int_{\Omega} g'(d) \left[ \mathcal{H}^{+} + (1 -b)p \diver \tensor{u} + \tensor{u} \cdot \grad p \right] \phi \: \dd V + \int_\Omega \dfrac{\mathcal{G}_{c}}{L} \left( d \phi + L^{2} \grad d \cdot \grad \phi \right) \: \dd V = 0, 
    \label{eq:damage-evolution-residual}  \\ 
   \mathcal{R}_{p}(\tensor{u},d,p) := & 
    \int_\Omega  \dfrac{\bar{\phi}(d)}{K_{f}} \left(p - p_{n} \right) \psi \: \dd V
    + \int_\Omega \bar{b}(d) \diver \left(\tensor{u} - \tensor{u}_{n} \right) \psi \: \dd V \nonumber \\ 
    & + \int_\Omega \Delta t \dfrac{k_{m}\tensor{1} + [1 - g(d)]\tensor{k}_{f} }{\mu_{f}} \cdot \left( \grad p - \rho \tensor{g}\right) \cdot \grad \psi \: \dd V \nonumber \\ 
    & - \int_\Omega Q \Delta t \psi \: \dd V + \int_{\pd_{q} \Omega} \Delta t\, \hat{q} \psi \: \dd A = 0. \label{eq:flow-eq-residual}
\end{align}
Here, $\Delta t = t_{n+1} - t_n$ denotes the time increment, 
and $(\cdot)_{n}$ denotes quantities at time step $t_{n}$. 
For notational brevity, quantities at time step $t_{n+1}$ have been written without any subscript. 
The rest of the finite element discretization of the above variational equations is straightforward and omitted for brevity.

For robust and efficient solution of the governing equations, we sequentially solve the hydromechanical problem, Eqs.~\eqref{eq:momentum-balance-residual} and~\eqref{eq:flow-eq-residual}, and the phase-field problem, Eq.~\eqref{eq:damage-evolution-residual}, extending the staggered solution algorithm proposed by~\cite{miehe2010phase}.
Each problem is solved by Newton's method.
When modeling propagating fractures, we perform at least 10 staggered iterations to ensure sufficient accuracy.
However, when modeling non-growing fractures, we just have to solve Eq.~\eqref{eq:damage-evolution-residual} once to initialize the distribution of the phase field.
Also, to solve the hydromechanical problem efficiently, we make use of stabilized equal-order linear finite elements~\citep{white2008stabilized} along with a block-preconditioned Newton--Krylov solver~\citep{white2011block}.
In what follows, we describe how to evaluate the crack opening, $\omega$, in Eq.~\eqref{eq:flow-eq-residual}, during the numerical solution procedure.

\subsection{Algorithm for crack opening calculation}

Algorithm~\ref{algo:aperture-update} presents the procedure to compute the crack opening at a material (quadrature) point.
In this algorithm, we first check if the material point is inside or outside the crack according to its phase-field value. 
If $d$ is smaller than a small threshold $d_\mathrm{tol}$ (\eg~$10^{-6}$), we treat that the material point is outside the crack and set $\omega=0$. 
Otherwise, we treat that the material point is inside the crack and calculate the crack density function $\Gamma_{d}(d,\grad d)$ based on the phase-field value and its gradient at the material point. 
In doing so, a small threshold $k_\mathrm{tol}$ is introduced to avoid the denominator in Eq.~\eqref{eq:omega-final} becoming zero.  
Finally, we update the crack opening, $\omega$ according to Eq.~\eqref{eq:omega-final}. 

For computational robustness, we call the algorithm after solving both the hydromechanical and phase-field problems.
In this way, $\omega$ is updated explicitly such that it does not need to be linearized during a Newton iteration.
As the time step size during fracture propagation is usually very small, this explicit update does not compromise the solution accuracy significantly.
If desired, however, one can update $\omega$ implicitly by calling the algorithm during the hydromechanical solution stage.
Also, the fracture normal vector, $\tensor{n}$, is pre-determined in this work, as the fracture orientations in all numerical examples presented below are clearly defined.
To model propagating fractures with complex geometry, one can use a common and simple approximation, $\tensor{n} \approx \grad d / \lvert\lvert \grad d \rvert\rvert$, or use more advanced and accurate algorithms in the literature (\eg~\cite{ziaei2016identifying,fei2020phase}).
\begin{algorithm}[htbp]
  \setstretch{1.3}
  \caption{Calculation of crack opening at a material point}
  \begin{algorithmic}[1]
  \Require $\tstrain$, $d$, $\grad d$, $p$, $\tensor{n}$, and material parameters 
  \If{$d < d_\mathrm{tol}$}
    \State Outside the crack, $\omega = 0$
  \Else
    \State Inside the crack.
    \State Calculate $\Gamma_{d}$ as Eq.~\eqref{eq:gamma_d}.  
    \State Let $\Gamma_{d} = \max \{\Gamma_{d}(d, \grad d), k_\mathrm{tol} \}$ with $k_\mathrm{tol} = 10^{-6}$. 
    \State Update $\omega = [(\lambda \tensor{1} + 2\mu \tensor{n} \dyadic \tensor{n}):\tstrain + p ]/[\Gamma_{d}(d, \grad d) (\lambda + 2\mu)]$.
  \EndIf
  \Ensure $\omega$. 
  \end{algorithmic}
  \label{algo:aperture-update}
\end{algorithm}

\section{Verification}
\label{sec:verification}

In this section, we verify the proposed crack-opening calculation method with benchmark examples in the literature.
Three examples are chosen: (i) the Sneddon problem~\citep{sneddon1946distribution}, for which the crack opening of a pressurized crack can be computed analytically, (ii) injection into a saturated porous medium with a fracture, for which numerical solutions from well-verified discrete methods are available~\citep{khoei2014mesh,cusini2021simulation}, and (iii) the Kristianovich–Geertsma–de Klerk (KGD) hydraulic fracture propagation, for which analytical solutions are available~\citep{detournay2003near}.
In all examples, we apply the proposed method with varied element sizes ($h$) and phase-field regularization lengths ($L$). 
The numerical results are produced using an in-house finite element code used in our previous work (\eg~\cite{choo2015stabilized,choo2018coupled,choo2019stabilized}), which is built on the \texttt{deal.II} library~\citep{arndt2021deal}.
In all cases, we use quadrilateral elements with linear shape functions, neglect body force, and assume plane-strain conditions. 
Also, we initialize phase-field fractures following the approach proposed by~\cite{borden2012phase} that prescribes high crack driving forces at the preexisting fractures and solves the phase-field governing equation~\eqref{eq:damage-eq-H+}.
For brevity, we omit the specific formulation for the initial crack driving force, referring to Appendix~A in~\cite{borden2012phase}.

\subsection{Sneddon problem}

\begin{figure}
    \centering
    \includegraphics[width=0.45\textwidth]{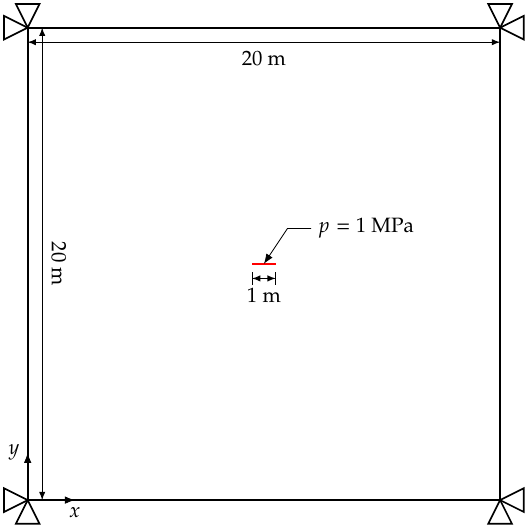}
    \caption{Sneddon problem: domain geometry and boundary conditions.}
    \label{fig:sneddon-setup}
\end{figure}

Our first example is the Sneddon problem~\citep{sneddon1946distribution}---a pressurized crack inside an infinite 2D elastic domain.
As a proxy of an infinite domain containing a crack, we consider a 20 m $\times$ 20 m square domain with a 1 m long horizontal fracture at the center, as depicted in Fig.~\ref{fig:sneddon-setup}.
A uniform pressure of 1 MPa is applied along the fracture, and all the external boundaries are subjected to zero displacement conditions.
Following the setup of the Sneddon problem, the domain is nonporous and impermeable.
In this case, we only have to solve the momentum balance equation~\eqref{eq:momentum-balance-residual} setting $p$ in the first term (matrix pressure) as zero and $p$ in the second term (fracture pressure) as 1 MPa, without solving the mass balance equation~\eqref{eq:flow-eq-residual}.
The analytical solution to the crack opening along the fracture is given by
\begin{align}
    \omega_\text{analytical}(x) = \dfrac{4pa(1 - \nu^2)}{E}\sqrt{1 - \dfrac{(x - x_\text{mid})^2}{a^2}} ,
\end{align}
where $x_\text{mid}$ is the $x$-coordinate of the crack center ($x_\text{mid}=10$ m in our case), $a$ is half of the fracture length ($a=0.5$ m in our case), and $E$ and $\nu$ are Young's modulus and Poisson's ratio, respectively. 
Here, we set $E = 1000$ MPa and $\nu = 0.15$. 

\begin{figure}
    \centering
    \includegraphics[width=1.0\textwidth]{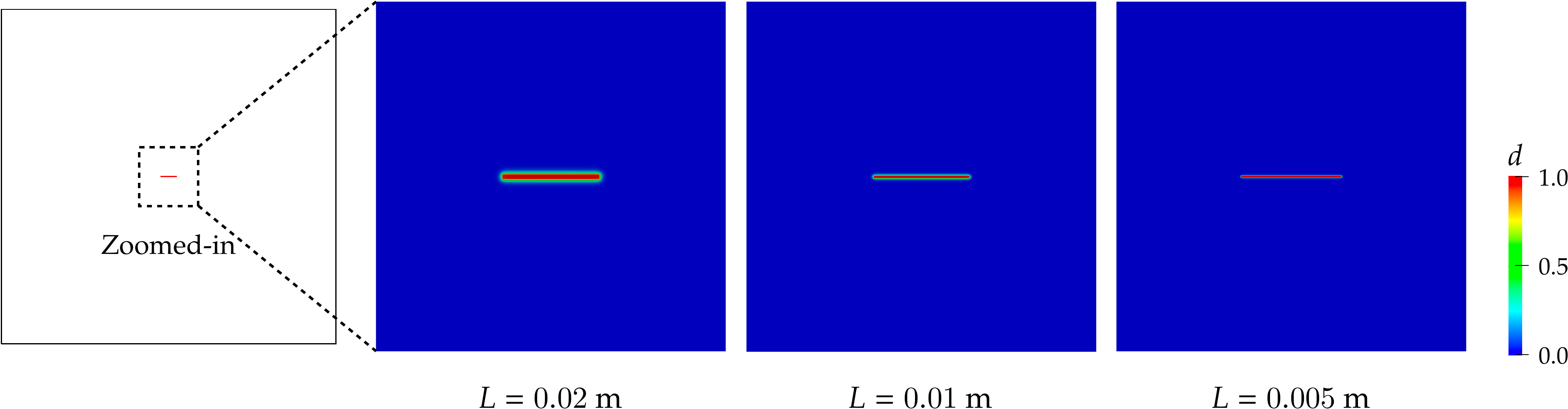}
    \caption{Sneddon problem: initial phase-field distribution with three different regularization lengths.}
    \label{fig:sneddon-pf}
\end{figure}

To examine the sensitivity of the proposed method to the phase-field regularization length, we repeatedly simulate the problem with three different values of $L$: 0.005 m, 0.01 m, and 0.02 m.
Figure~\ref{fig:sneddon-pf} shows the phase-field fractures created with these three values of $L$. 
Also, to assess mesh sensitivity, we repeat the calculation with three levels of discretization: $L/h=$ 5, 10, and 20. ($h$ denotes the element size.)
For computational efficiency, such a refinement is locally applied to the phase-field region.
Further, to evaluate the sensitivity of the proposed method to the element alignment, we also consider meshes with three different element orientations, $\theta=0^\circ$, $33.7^\circ$, and $45^\circ$, where $\theta$ is the orientation of elements from the horizontal. 
Unless otherwise noted, we present numerical results from meshes with $\theta=0^\circ$ (structured meshes aligned with the crack) in the following.

\begin{figure}[htbp]
    \centering
    \subfloat[]{\includegraphics[width=0.48\textwidth]{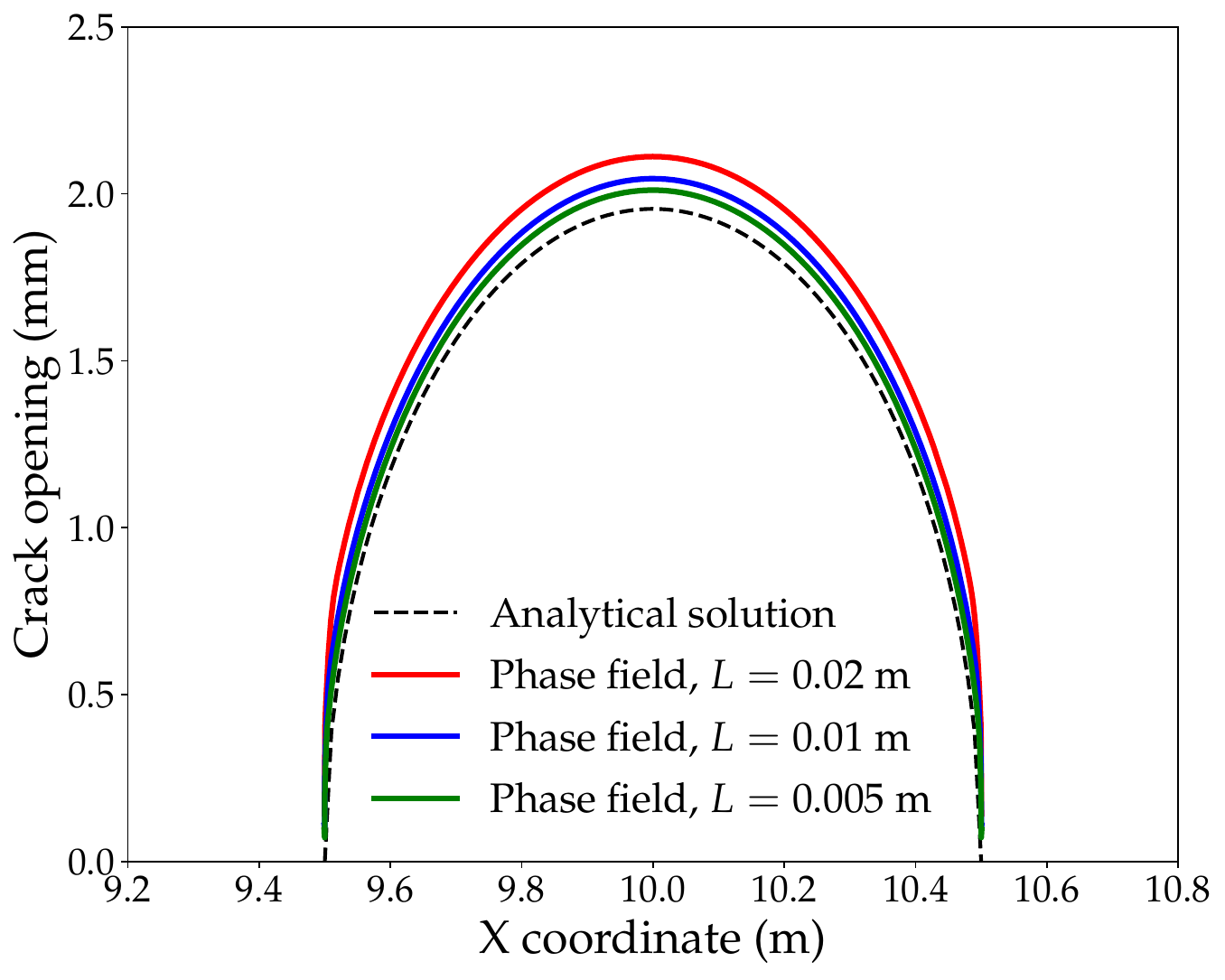}\label{fig:sneddon-results-length}}
    \hspace{0.02\textwidth}
    \subfloat[]{\includegraphics[width=0.48\textwidth]{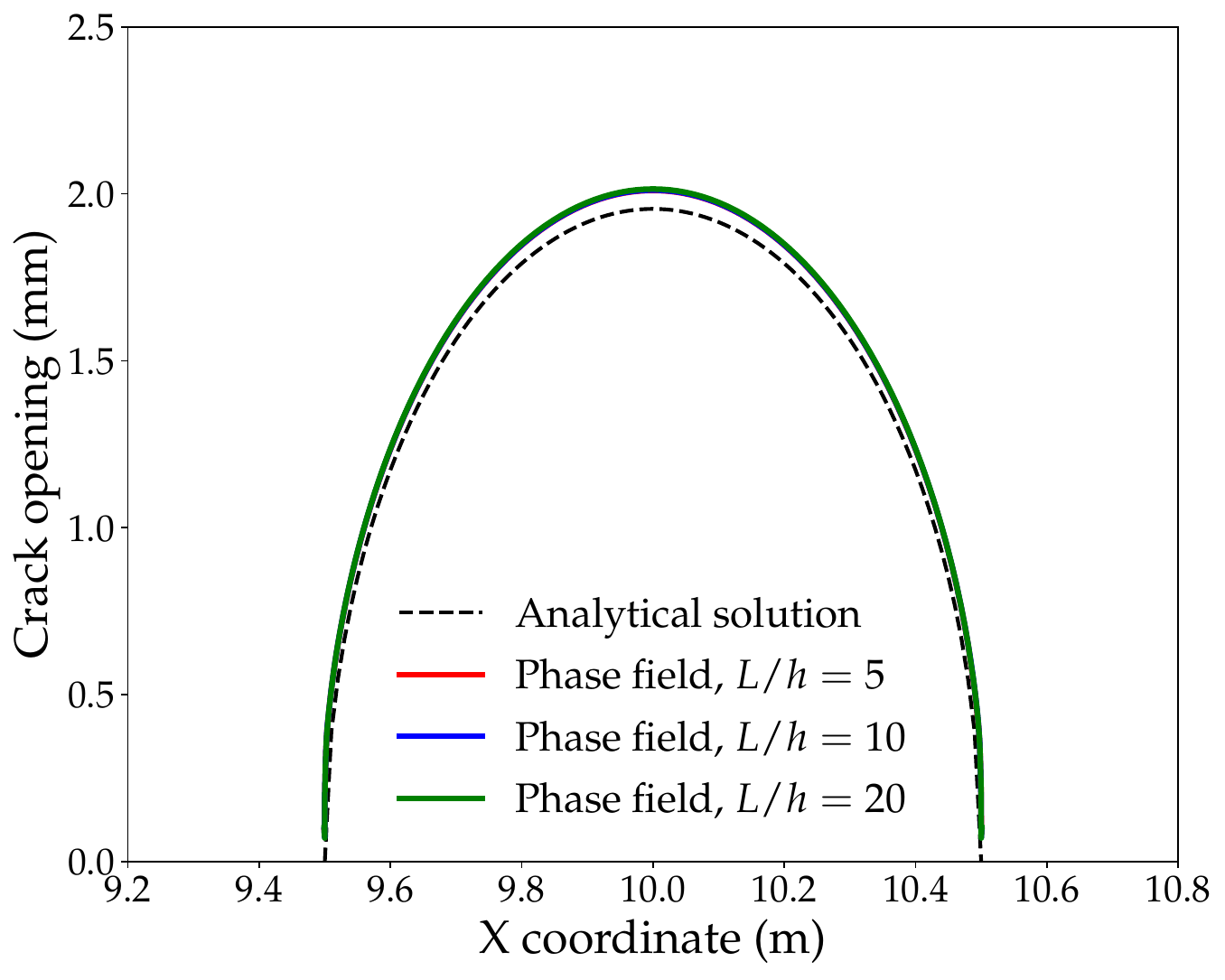} \label{fig:sneddon-results-mesh}} 
    \caption{Sneddon problem: comparison between the analytical solution and phase-field results with (a) different regularization lengths $L$ under a fixed refinement level as $L/h = 10$, and (b) different element sizes $h$ under a fixed regularization length as $L = 0.005$ m.}
\end{figure}

Figure~\ref{fig:sneddon-results-length} plots the crack openings calculated by the proposed method, with different regularization lengths under a fixed discretization level of $L/h=10$. 
The crack openings shown here are extracted from the material points closest to the centerline of the crack.
As can be seen, the calculated crack openings converge to the analytical solution as the regularization length becomes smaller. 
In Fig.~\ref{fig:sneddon-results-mesh}, we show how the calculated crack openings vary with element sizes under a fixed regularization length $L = 0.005$ m. 
The calculated crack openings also converge with mesh refinement.
It is worth noting the results obtained with $L/h=5$---nearly the coarsest level of discretization in phase-field modeling of fracture---are practically identical to those obtained with finer discretizations, namely, $L/h=10$ and 20.
This observation suggests that the proposed method can provide a fairly accurate result as long as the mesh is sufficiently fine for phase-field modeling.
Also notably, the numerical results confirm that the proposed method allows for the use of any element size regardless of the crack opening size.
This is not the case for the existing strain-based calculations where the element size cannot be smaller than the crack opening size.

\begin{figure}[htbp]
    \centering
    \subfloat[]{\includegraphics[width=\textwidth]{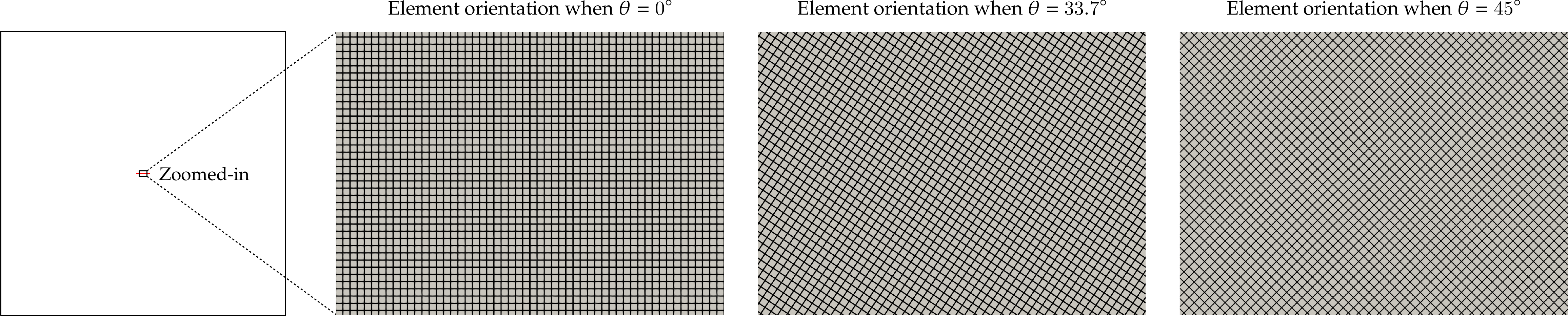}
    \label{fig:sneddon-unstructured-mesh}}
    \hspace{0.02\textwidth}
    \subfloat[]{\includegraphics[width=0.48\textwidth]{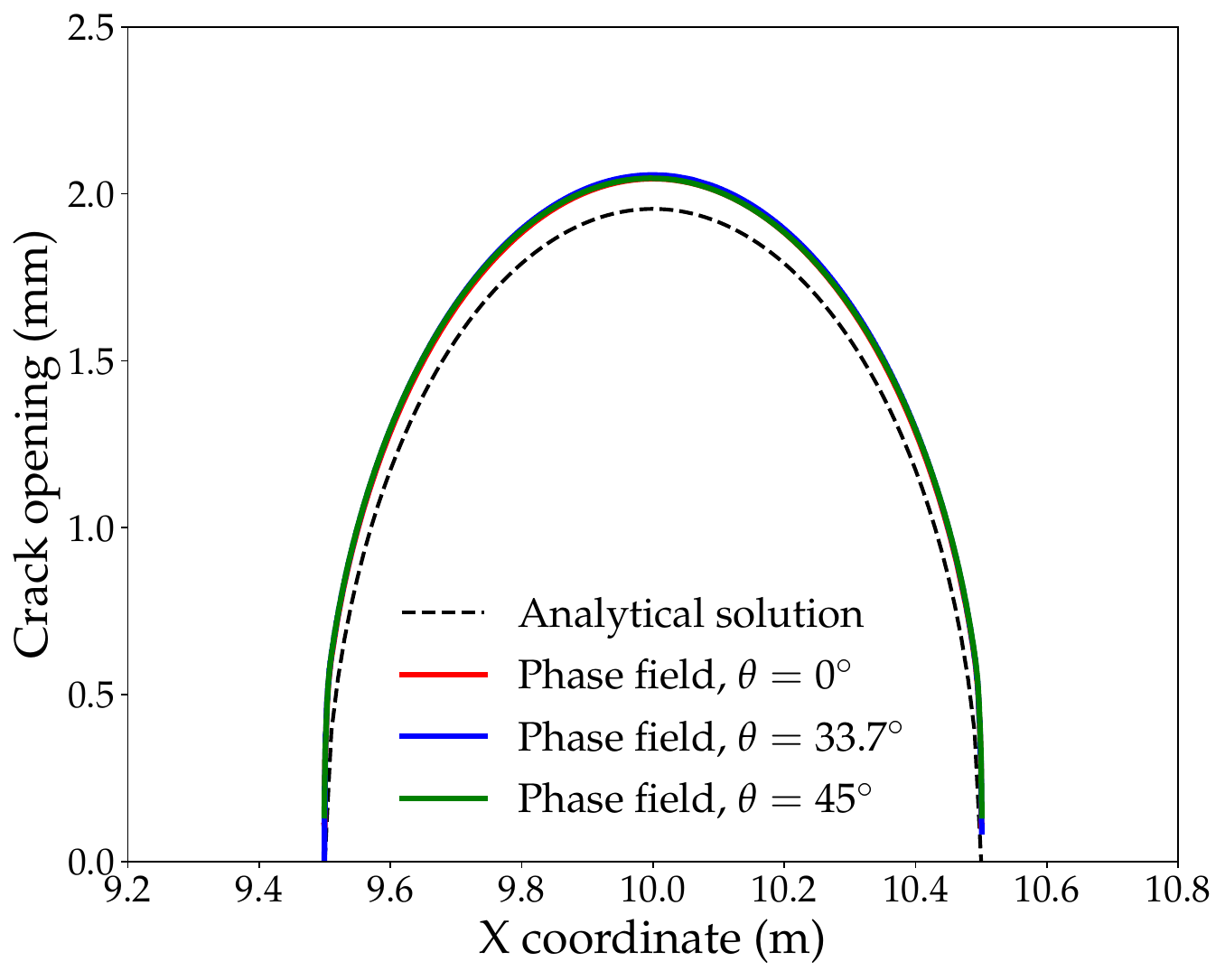} \label{fig:sneddon-results-unstructMesh}} 
    \caption{Sneddon problem: (a) meshes with different element orientation angles $\theta$, and (b) comparison between the analytical solution and phase-field results with different element orientations. (The refinement level and the regularization length are fixed as $L/h = 10$ and $L = 0.01$ m, respectively.)}
    \label{fig:sneddon-mesh-alignment}
\end{figure}

We further evaluate the sensitivity of the proposed method to the element alignment, by repeating the same problem with meshes depicted in Fig.~\ref{fig:sneddon-unstructured-mesh}.
Note that when $\theta$ is nonzero, the elements are not aligned with the crack orientation.
Figure~\ref{fig:sneddon-results-unstructMesh} compares the phase-field results obtained with the different element orientations along with the analytical solution.
One can see that the calculated crack openings are insensitive to the alignment of elements relative to the crack.
So the proposed method is verified to be insensitive not only to the element size but also to the element alignment.  
This feature distinguishes the proposed method from the existing strain-based methods in that the existing ones require elements to be aligned with the crack or an additional technique to correct errors associated with the misalignment of elements (see, \eg~Appendix B of~\cite{wilson2016phase}).

\subsection{Injection into a saturated porous medium with a fracture}

\begin{figure}
    \centering
    \includegraphics[width=0.5\textwidth]{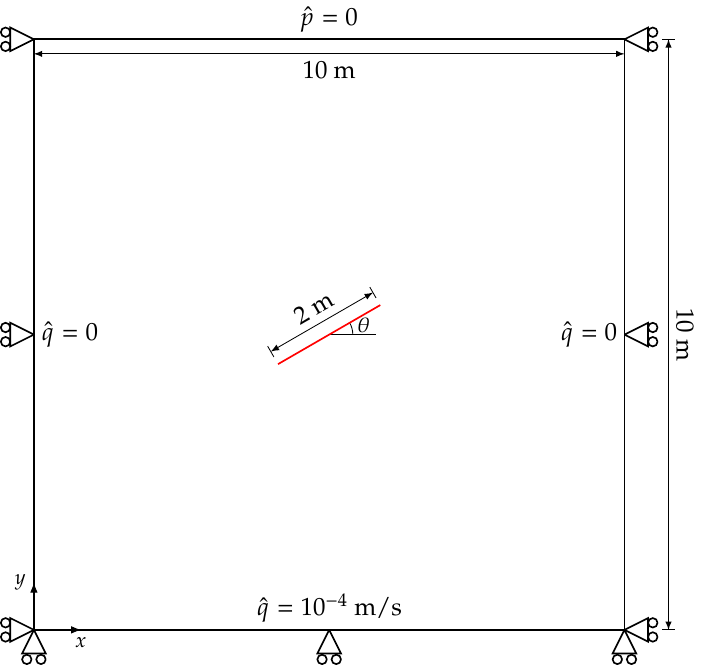}
    \caption{Injection into a saturated porous medium with a fracture: domain geometry and boundary conditions.}
    \label{fig:injectionSat-setup}
\end{figure}

In our second example, we apply the proposed method to a benchmark problem that has been simulated by two popular discrete methods for fracture, namely, the extended finite element method (XFEM)~\citep{khoei2014mesh} 
and the embedded finite element method (EFEM)~\citep{cusini2021simulation}. 
The problem considers an injection of fluid into a saturated porous medium containing a fluid-filled fracture.
As depicted in Fig.~\ref{fig:injectionSat-setup}, the problem domain is a 10 m square that possesses a 2 m long fracture with an inclination angle, $\theta$, at the center.
As for the displacement boundary conditions, all the boundaries except the top surface are supported by rollers. 
The flow boundary conditions are as follows: The top surface is a zero pressure boundary, the two lateral surfaces are impermeable, and the bottom surface is subjected to an influx with a constant rate of $q = 10^{-4}$ m/s. 
The total injection time is 10 s, and the time domain is discretized into 75 uniform time steps. 
The material properties of the domain are summarized in Table~\ref{tab:material-parameters-injSat}. 
\begin{table}[htbp]
    \centering
    \begin{tabular}{l|c|c|c}
    \hline 
      \textbf{Parameter}   & \textbf{Symbol} & \textbf{Unit} & \textbf{Value}  \\
    \hline
      Young's modulus & $E$ & GPa & 9.0 \\ 
      Poisson's ratio & $\nu$ & - & 0.4 \\ 
      Biot coefficient & $b$ & - & 1.0 \\
      Matrix permeability & $k_{m}$ & m$^2$ & $10^{-12}$ \\ 
      Porosity & $\phi_{m}$ & - & 0.3 \\ 
      Fluid viscosity & $\mu_{f}$ & Pa$\cdot$s & $10^{-3}$ \\ 
      \hline
    \end{tabular}
    \caption{Injection into a saturated porous medium with a fracture: material properties.}
    \label{tab:material-parameters-injSat}
\end{table}

For a thorough verification, we consider three cases with varied fracture inclination angles: (i) $\theta = 30^\circ$, (ii) $\theta = 45^\circ$, and (iii) $\theta=60^\circ$. 
For the $\theta=30^\circ$ and $45^\circ$ cases, we repeat the phase-field simulations with three regularization lengths, namely, $L = 0.015$ m, 0.02 m, and 0.03 m. 
Also, similar to the previous problem, we discretize the domain by a structured mesh in which the elements around the fracture are locally refined with three levels of discretization, namely, $L/h=5$, 10, and 20. 

\begin{figure}[htbp]
    \centering
    \includegraphics[width=0.8\textwidth]{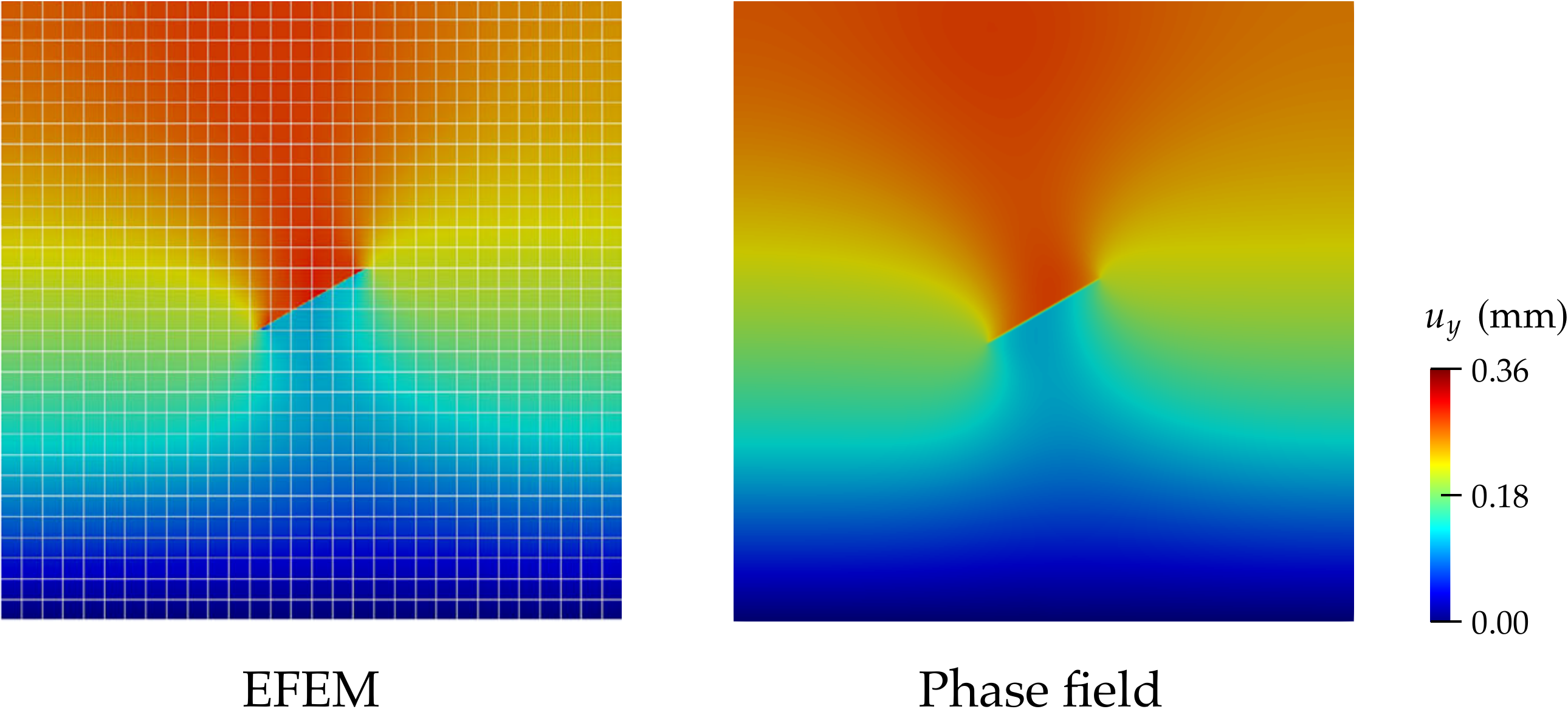}
    \caption{Injection into a saturated porous medium with a fracture: comparison between the phase-field and EFEM solutions~\citep{cusini2021simulation} in terms of the final $y$-displacement field when $\theta = 30^\circ$.}
    \label{fig:injectionSat-30deg-compare}
\end{figure}

\begin{figure}[htbp]
    \centering
    \includegraphics[width=0.8\textwidth]{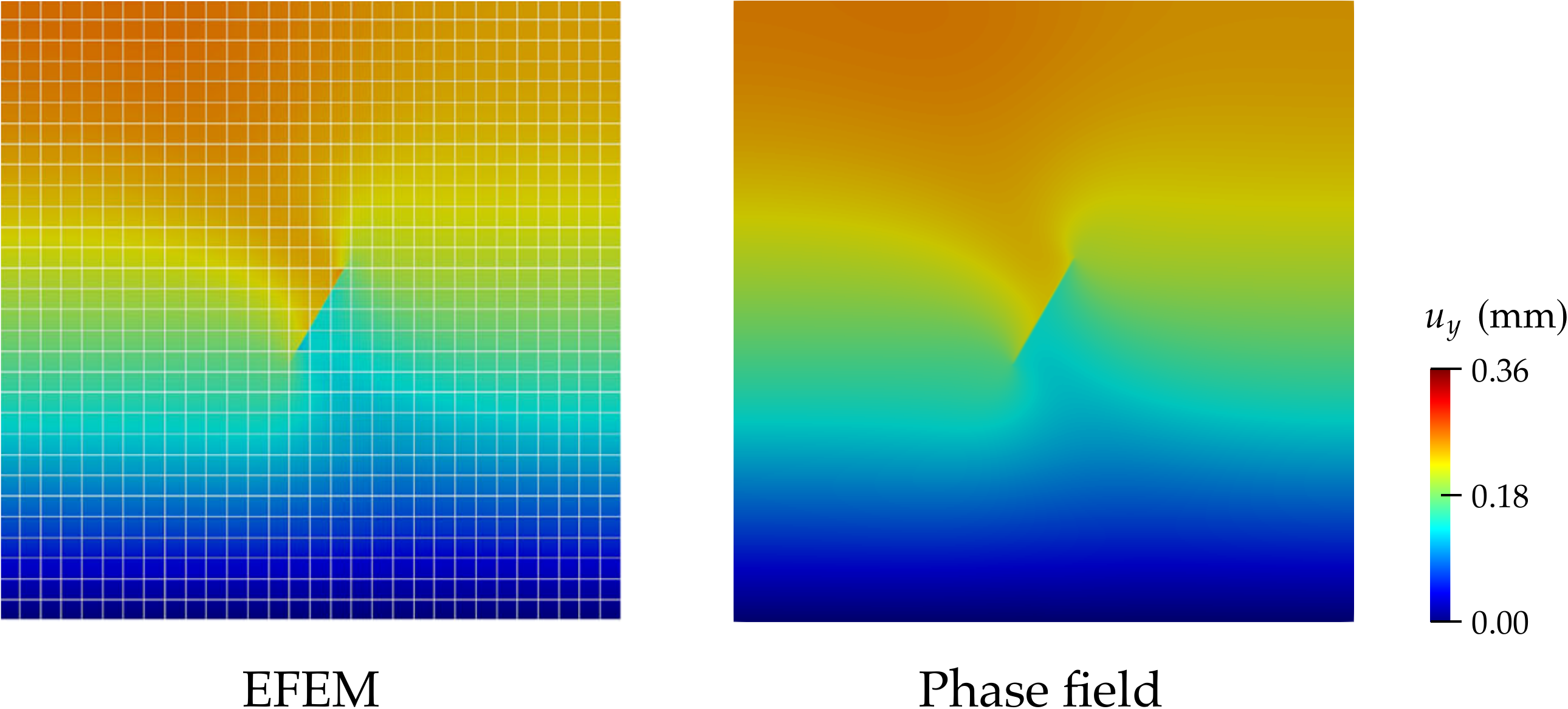}
    \caption{Injection into a saturated porous medium with a fracture: comparison between the phase-field and EFEM solutions~\citep{cusini2021simulation} in terms of the final $y$-displacement field when $\theta = 60^\circ$.}
    \label{fig:injectionSat-60deg-compare}
\end{figure}

We first compare the phase-field solutions with the EFEM solutions in~\cite{cusini2021simulation} in terms of the $y$-displacement field at the end of injection.
The phase-field solutions are obtained with the regularization length $L = 0.02$ m and a mesh satisfying $L/h=10$ around the phase-field fracture. 
Figure~\ref{fig:injectionSat-30deg-compare} presents the comparison when $\theta = 30^\circ$. 
We can see that the phase-field and EFEM solutions are virtually identical, clearly exhibiting the influence of the fracture at the domain center.
Figure~\ref{fig:injectionSat-60deg-compare} shows the same comparison when $\theta$ is changed to $60^\circ$.
Again, the phase-field and EFEM solutions are nearly the same, although the displacement fields have been changed with an increase in the inclination angle. 
Since the displacement field is coupled with the fluid pressure field, the agreement between the phase-field and EFEM solutions indicates indirectly that the fracture permeability field---governed by the crack opening displacement field---is correctly calculated by the proposed method.

\begin{figure}
    \centering
    \includegraphics[width=0.5\textwidth]{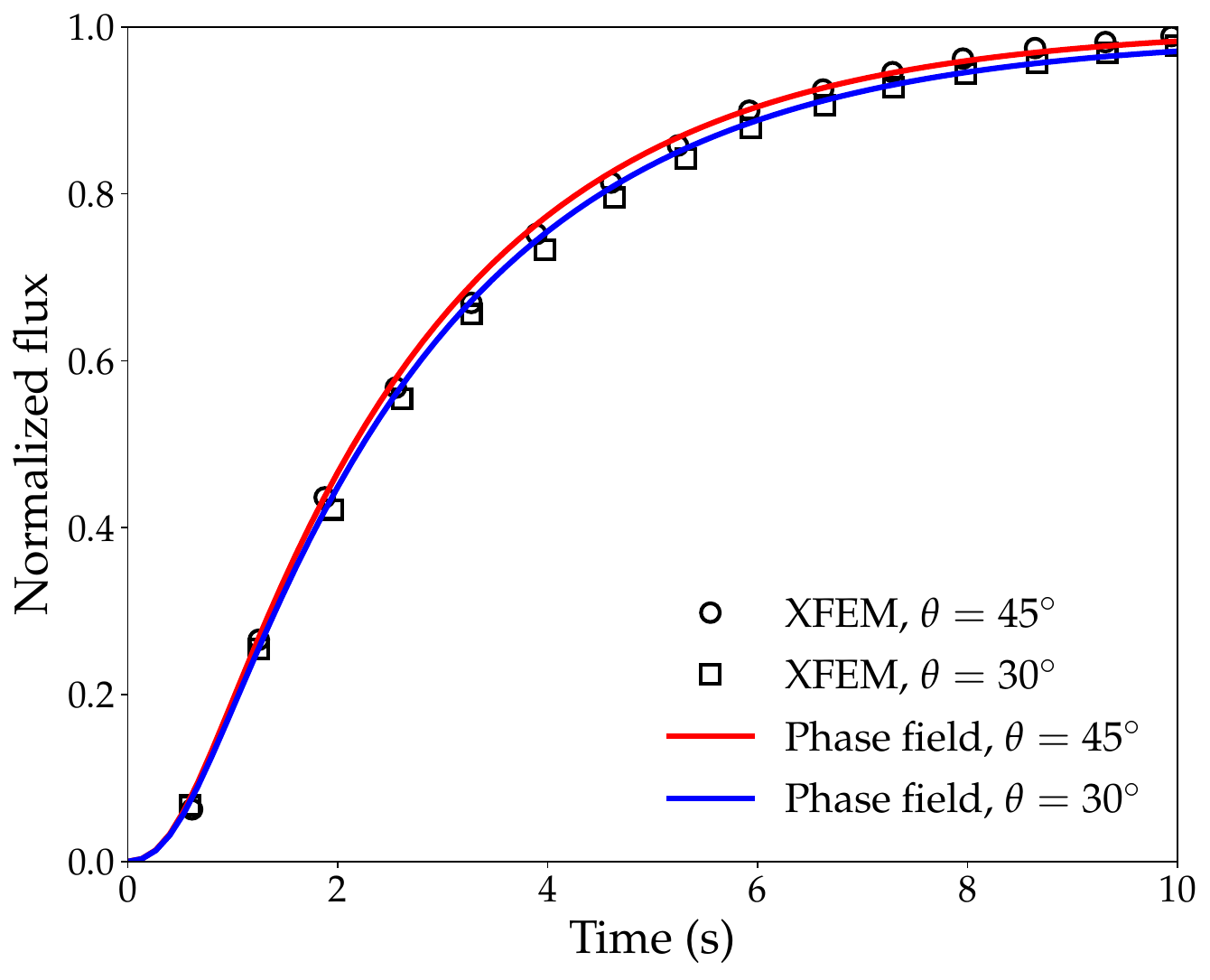}
    \caption{Injection into a saturated porous medium with a fracture: comparison between the phase-field and XFEM solutions~\citep{khoei2014mesh} in terms of the time evolution of the normalized flux at the top boundary when $\theta = 30^\circ$ and $45^\circ$.}
    \label{fig:injectionSat-30deg-compareFlux}
\end{figure}

Next, we compare the phase-field solutions with the XFEM solutions in~\cite{khoei2014mesh} in terms of the time evolution of the normalized fluxes at the top boundary. Figure~\ref{fig:injectionSat-30deg-compareFlux} shows the comparison when $\theta = 30^\circ$ and $45^\circ$.
As can be seen, for both inclination angles, the phase-field results are very close to the XFEM solutions during the entire injection period. 
It can also be confirmed that the phase-field model correctly captures the trend that the flux on the top boundary becomes faster with an increase in $\theta$. 
Since the flux solution must be directly controlled by the permeability of the fracture, it can be confirmed that the proposed method has calculated the crack opening correctly.

\begin{figure}[htbp]
    \centering
    \subfloat[]{\includegraphics[width=0.47\textwidth]{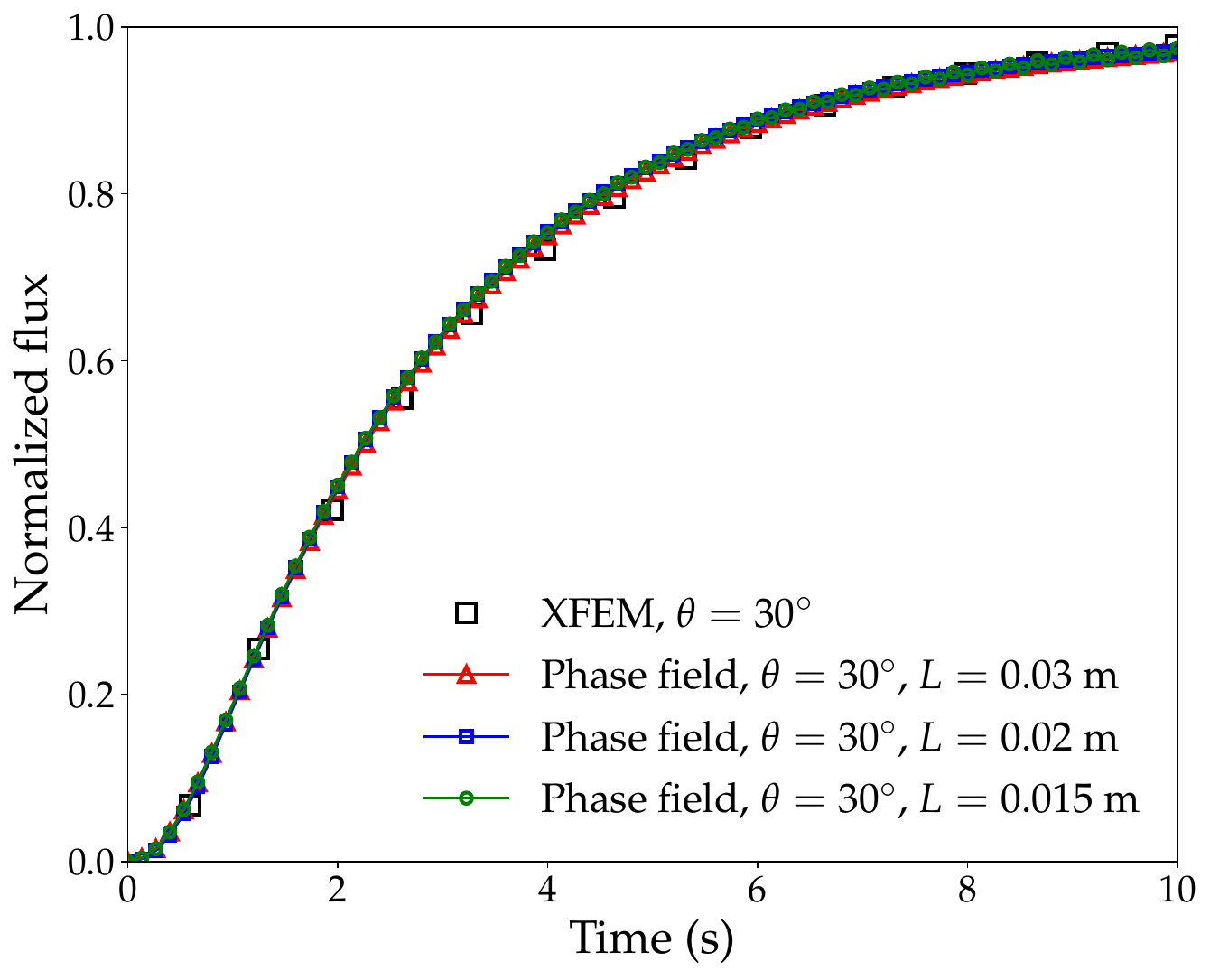} \label{fig:injectionSat-lengthConvergence-30deg}}
    \hspace{0.02\textwidth}
    \subfloat[]{\includegraphics[width=0.47\textwidth]{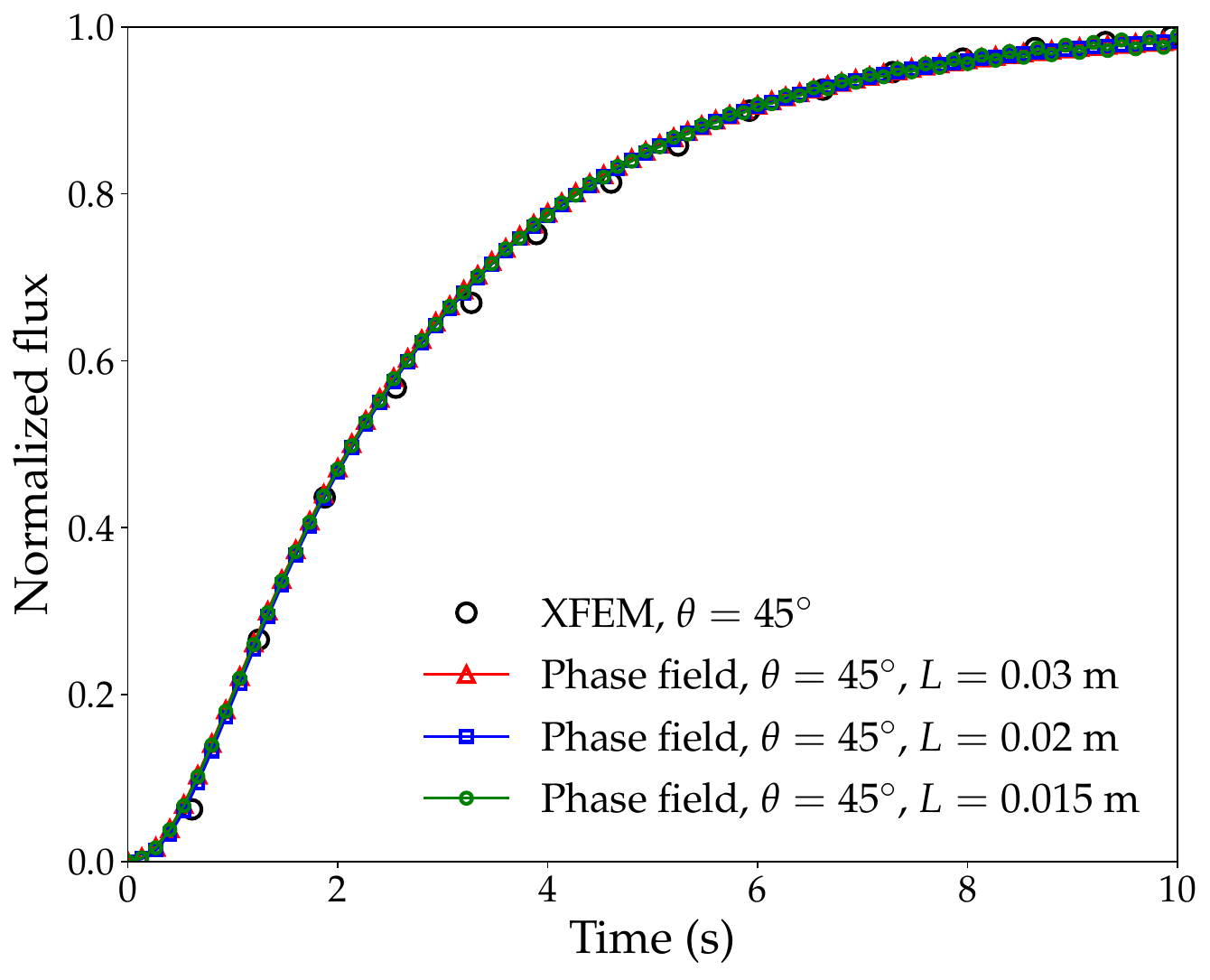} \label{fig:injectionSat-lengthConvergence-45deg}}
    \caption{Injection into a saturated porous medium with a fracture: sensitivity of the phase-field solutions to the regularization length, $L$: (a) $\theta=30^\circ$ and (b) $\theta = 45^\circ$. 
    The discretization level is $L/h = 10$ in all the cases.}
    \label{fig:injectionSat-sensitivity-30deg}
\end{figure}

\begin{figure}[htbp]
    \centering
    \subfloat[]{\includegraphics[width=0.47\textwidth]{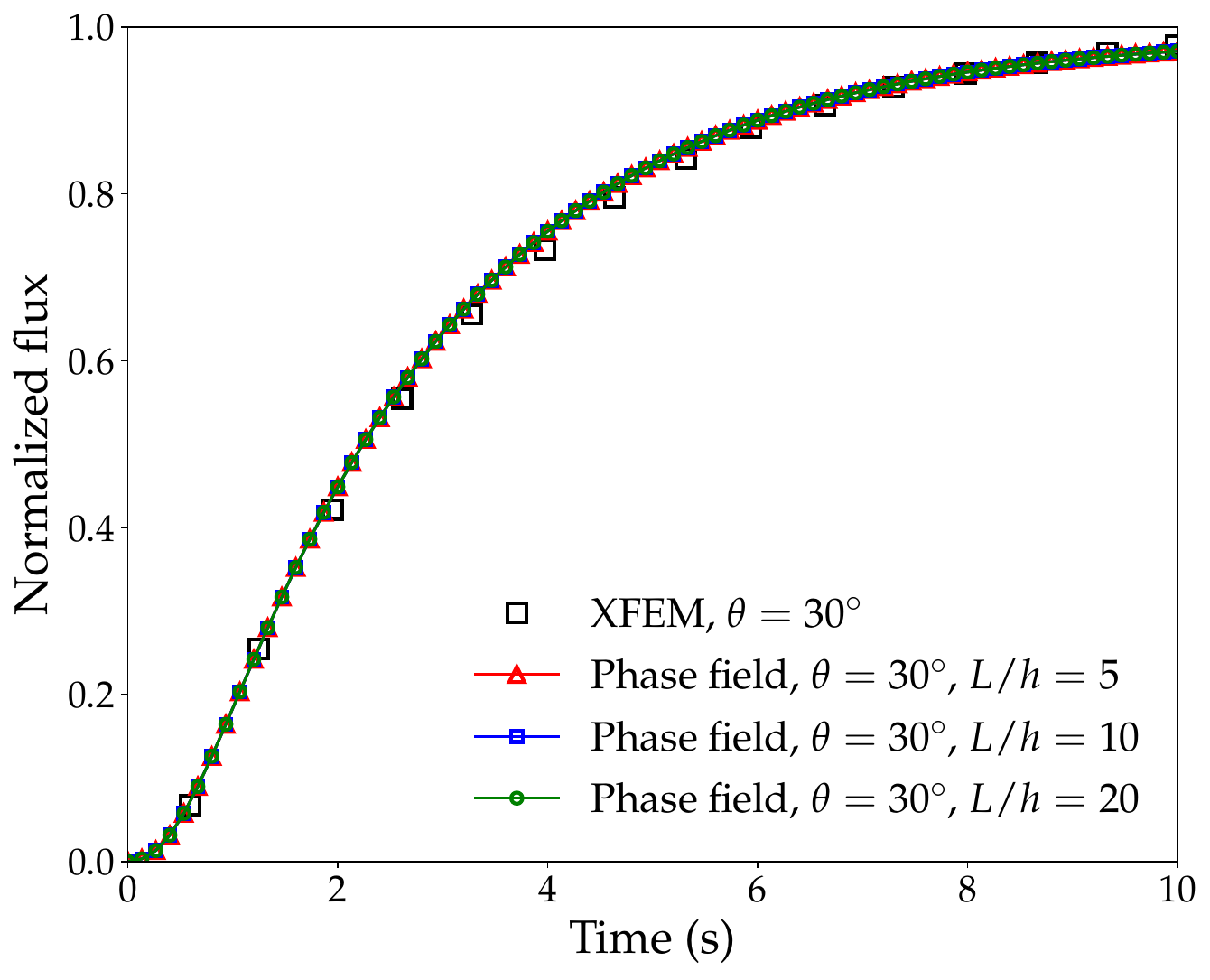} \label{fig:injectionSat-meshConvergence-30deg}}
    \hspace{0.02\textwidth}
    \subfloat[]{\includegraphics[width=0.47\textwidth]{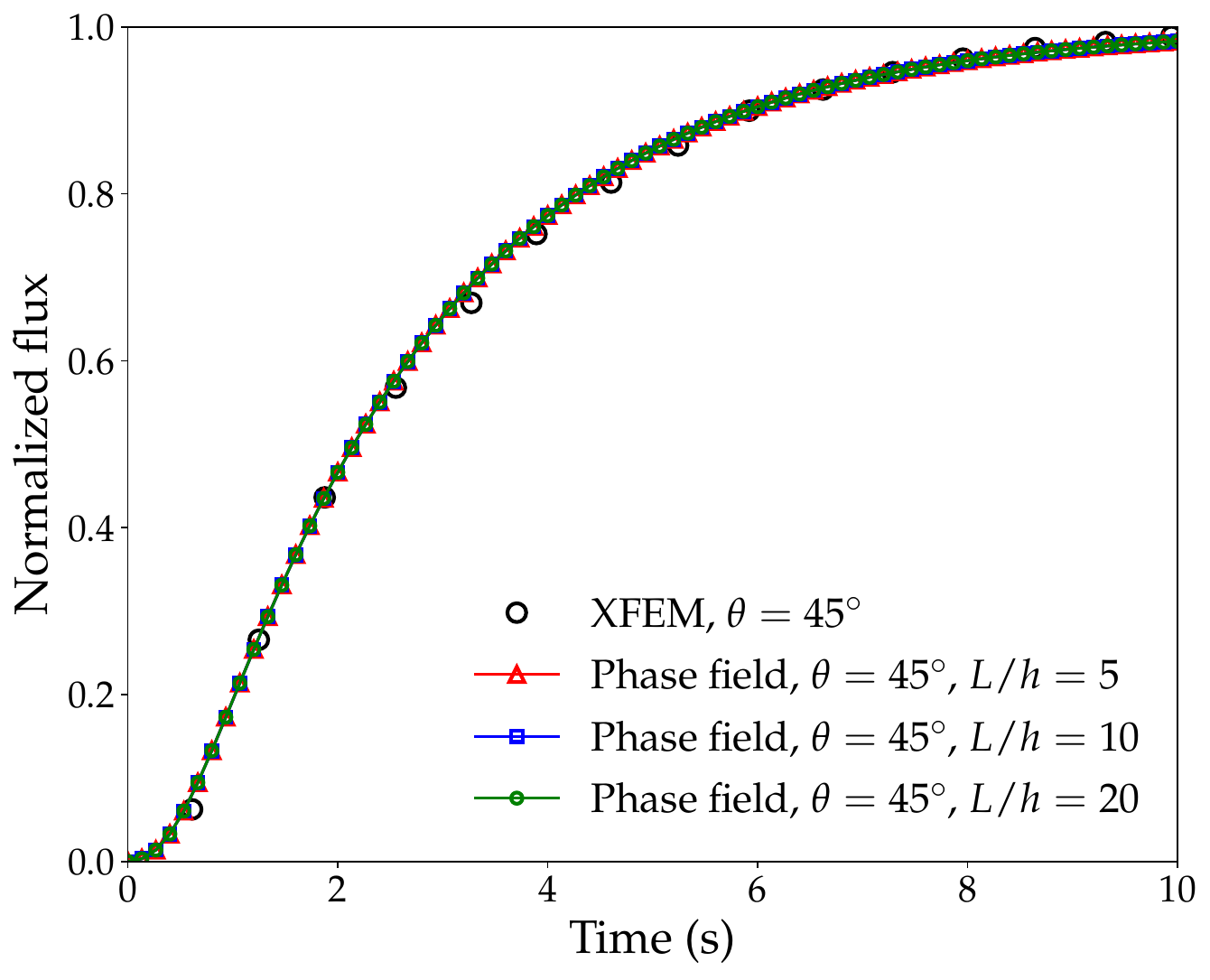} \label{fig:injectionSat-meshConvergence-45deg}}
    \caption{Injection into a saturated porous medium with a fracture: sensitivity of the phase-field solutions to the element size, $h$: (a) $\theta=30^\circ$ and (b) $\theta = 45^\circ$. The regularization length is $L = 0.02$ m in all the cases.}
    \label{fig:injectionSat-sensitivity-45deg}
\end{figure}

Lastly, we check the sensitivity of the phase-field solutions to the regularization length and the element size.
Figures~\ref{fig:injectionSat-lengthConvergence-30deg} and \ref{fig:injectionSat-lengthConvergence-45deg} show the flux solutions obtained with three different regularization lengths for the cases of $\theta = 30^\circ$ and $45^\circ$, respectively. 
Likewise, in Figs.~\ref{fig:injectionSat-meshConvergence-30deg} and \ref{fig:injectionSat-meshConvergence-45deg}, we show the flux solutions obtained with three different discretization levels for the cases of $\theta = 30^\circ$ and $45^\circ$, respectively.
All the results consistently indicate that the phase-field solutions, in which the proposed method for crack opening calculation plays a critical role, are insensitive to the regularization length and the element size.

\subsection{KGD hydraulic fracture propagation}
Our third and last example is intended to verify the proposed method for a propagating fracture.
For this purpose, we consider the KGD problem---a hydraulic fracture propagating in an infinite, elastic solid under plane strain conditions---for which analytical solutions are available for crack opening and other related quantities. 
To evaluate the analytical solution of the problem, we first introduce a dimensionless viscosity $\mathcal{M}$, defined as~\citep{garagash2006plane} 
\begin{align}
    \mathcal{M} := \dfrac{\mu'Q}{E'}\left(\dfrac{E'}{K'} \right)^{4},
\end{align} 
where $\mu' := 12 \mu$, $E' := E/(1 - \nu^2)$, and $K' := \sqrt{32\mathcal{G}_{c}E'/\pi}$. 
If $\mathcal{M}$ is smaller than $\mathcal{M}_{0} = 3.4 \times 10^{-3}$, the fracture propagation is in the toughness-dominated regime~\citep{garagash2006plane}.
In the toughness-dominated regime, the analytical solutions for the crack opening at the crack center ($\omega_{cc}$), net pressure at the crack center ($p_{cc}$), and the half crack length ($a$) are given by
\begin{align}
    \omega_{cc} (t) &= \dfrac{4p_{f}a}{E'}, \\ 
    p_{cc}(t) &= \left(\dfrac{2E'\mathcal{G}^2_{c}}{\pi Q t} \right)^{1/3}, \\
    a(t) &= \left(\dfrac{E'Q^{2}t^2}{4\pi \mathcal{G}_{c}} \right)^{1/3}.
\end{align}

\begin{figure}[htbp]
    \centering
    \includegraphics[width=0.45\textwidth]{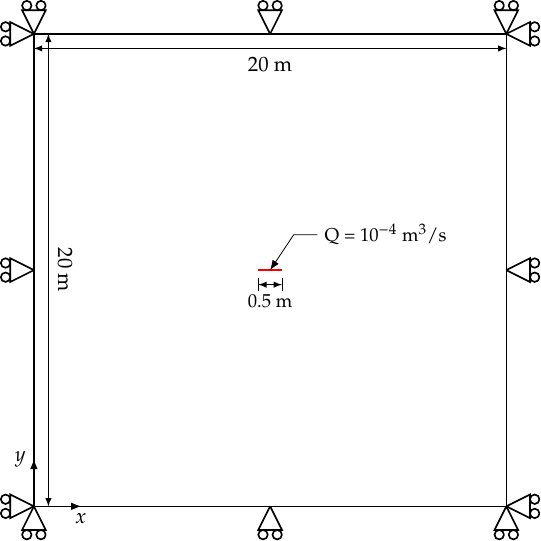} 
    \caption{KGD hydraulic fracture propagation: domain geometry and boundary conditions.}
    \label{fig:kgd-setup}
\end{figure}

For phase-field modeling of the KGD problem, we consider the setup depicted in Fig.~\ref{fig:kgd-setup}, in which a short (0.5 m long) initial crack is accommodated in a large (20 m by 20 m square) domain.
The fluid is injected into the center of the crack with a constant injection rate of $Q = 10^{-4}$ m$^{3}$/s.
The material properties adopted in this example are listed in Table~\ref{tab:material-parameters-kgd}. 
Note that we use a zero Biot coefficient and very low porosity and permeability values to emulate the nonporous and impermeable solid considered in the KGD problem. 
These material properties result in $\mathcal{M} = 4.8 \times 10^{-8}$, such that the fracture propagation is in the toughness-dominated regime.

\begin{table}[htbp]
    \centering
    \begin{tabular}{l|c|c|c}
    \hline 
      \textbf{Parameter}   & \textbf{Symbol} & \textbf{Unit} & \textbf{Value}  \\
    \hline
      Young's modulus & $E$ & GPa & 17.0 \\ 
      Poisson's ratio & $\nu$ & - & 0.15 \\ 
      Biot coefficient & $b$ & - & 0.0\\
      Critical fracture energy & $\mathcal{G}_{c}$ & J/m$^2$ & 200 \\
      Matrix permeability & $k_{m}$ & m$^2$ & $10^{-20}$ \\ 
      Porosity & $\phi_{m}$ & - & 0.01 \\ 
      Fluid viscosity & $\mu_{f}$ & Pa$\cdot$s & $10^{-8}$ \\ 
      Fluid bulk modulus & $K_{f}$ & GPa & 10 \\
      \hline
    \end{tabular}
    \caption{KGD hydraulic fracture propagation: material properties.}
    \label{tab:material-parameters-kgd}
\end{table}

As in the previous two examples, we repeat the simulation with three different phase-field regularization lengths, namely $L = 0.0025$ m, 0.005 m, and 0.01 m, to assess the length sensitivity of the phase-field solutions. 
Also, to examine the mesh sensitivity, we repeat the same problem with three different discretization levels, namely $L/h = 5$, 10, and 20, locally refining the potential fracture propagation path.   
We set the total injection time to be 4 s and discretize it into 250 uniform time steps.

\begin{figure}[htbp]
    \centering
    \subfloat[]{\includegraphics[width=0.45\textwidth]{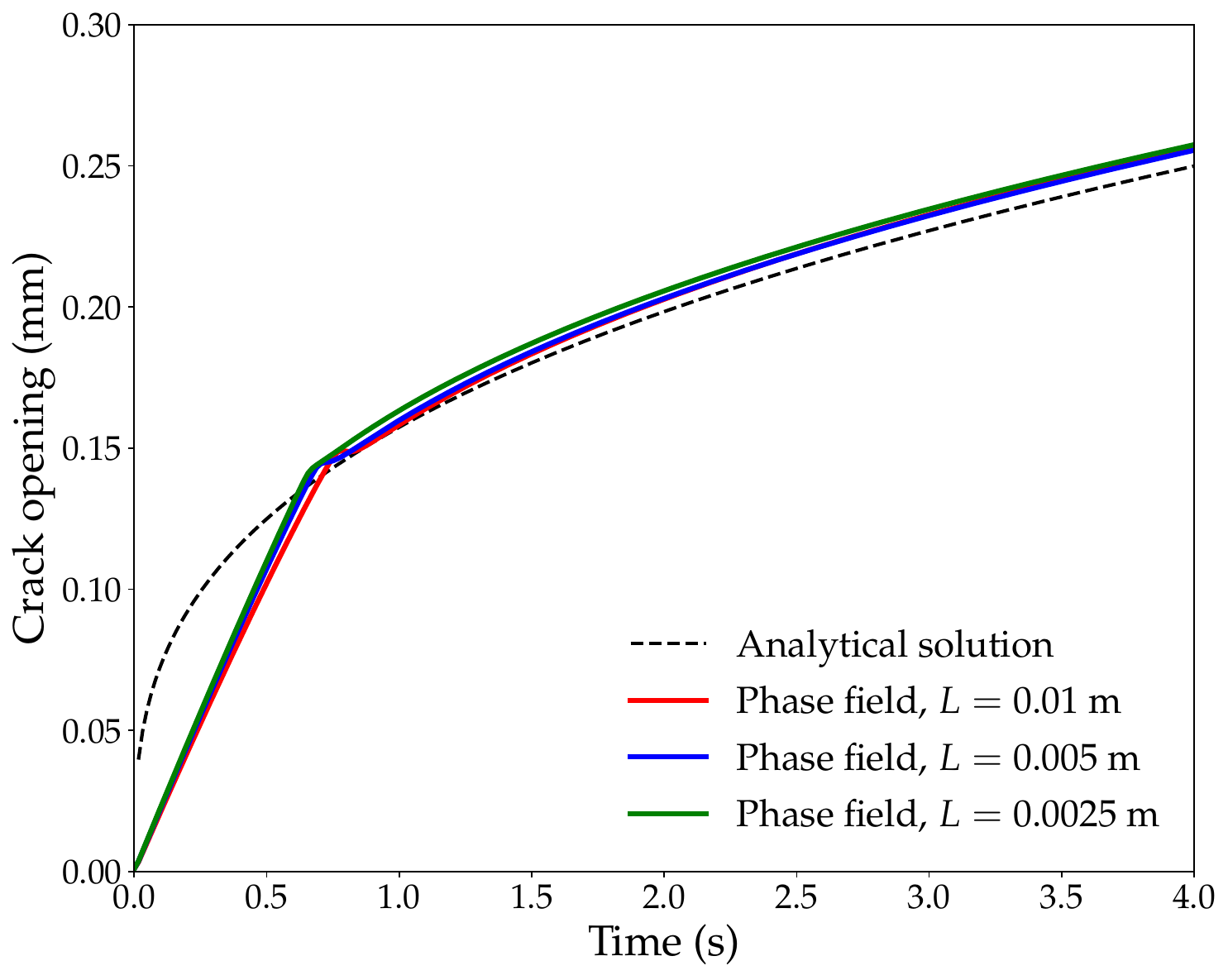} \label{fig:kgd-aperture-lengthCompare}}\\
    \subfloat[]{\includegraphics[width=0.45\textwidth]{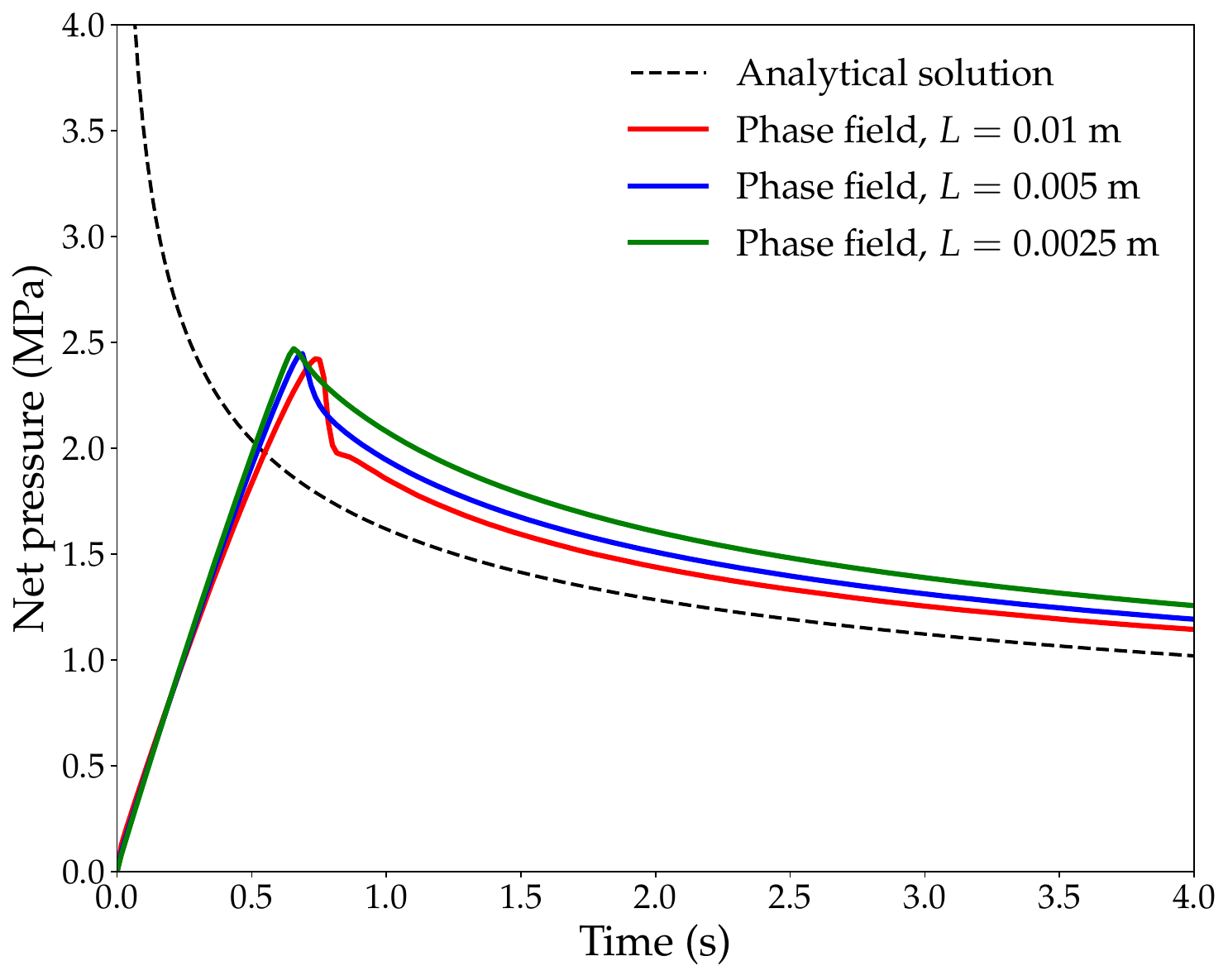} \label{fig:kgd-pressure-lengthCompare}} \\
    \subfloat[]{\includegraphics[width=0.45\textwidth]{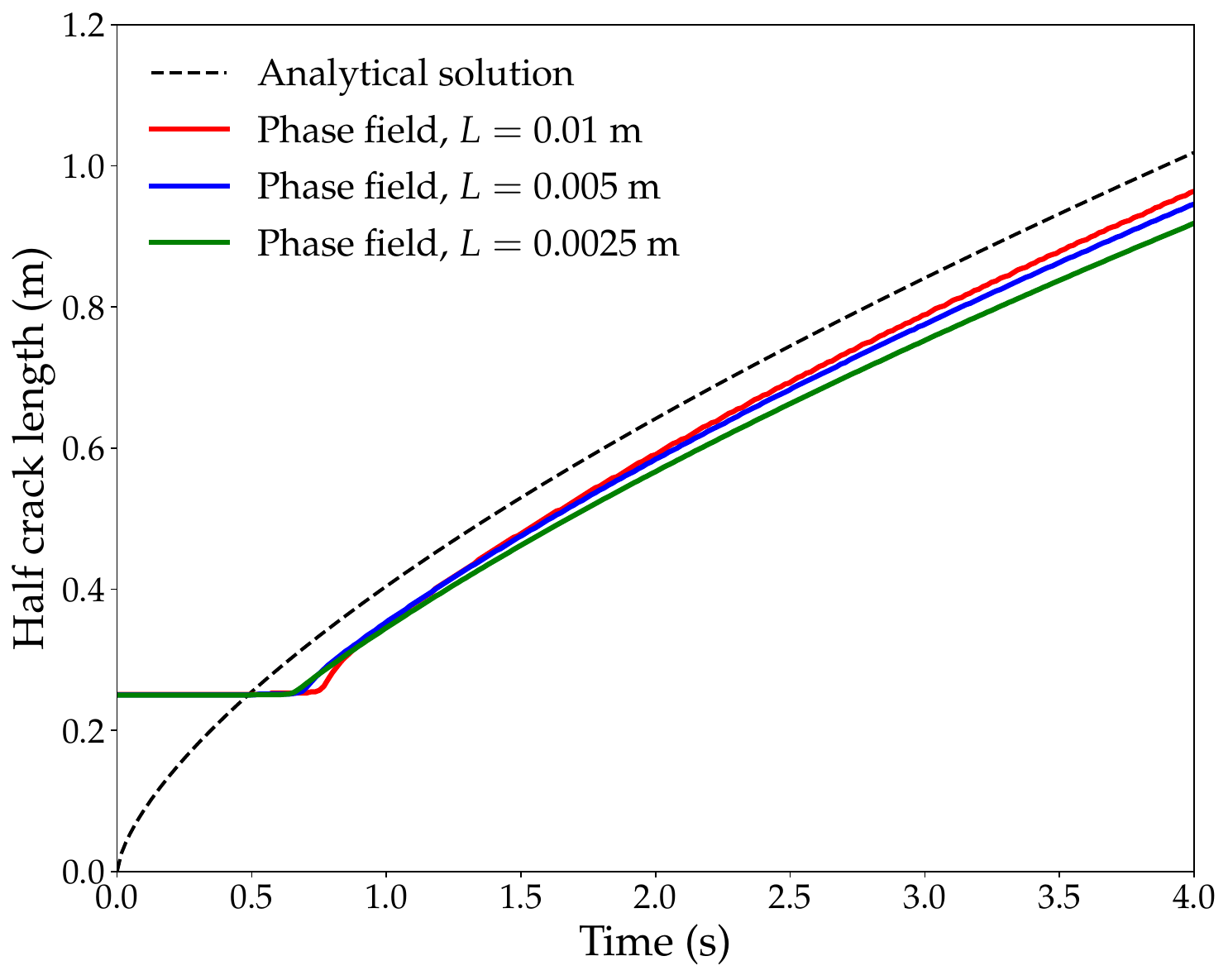} \label{fig:kgd-crackLength-lengthCompare}}
    \caption{KGD hydraulic fracture propagation: comparison between the phase-field and analytical solutions in terms of the (a) crack opening at the center (b) net pressure at the crack center, and (c) half crack length, with different regularization lengths $L$. The discretization level is $L/h = 5$ in all the cases.}
    \label{fig:kgd-lengthCompare}
\end{figure}

\begin{figure}[htbp]
    \centering
    \subfloat[]{\includegraphics[width=0.45\textwidth]{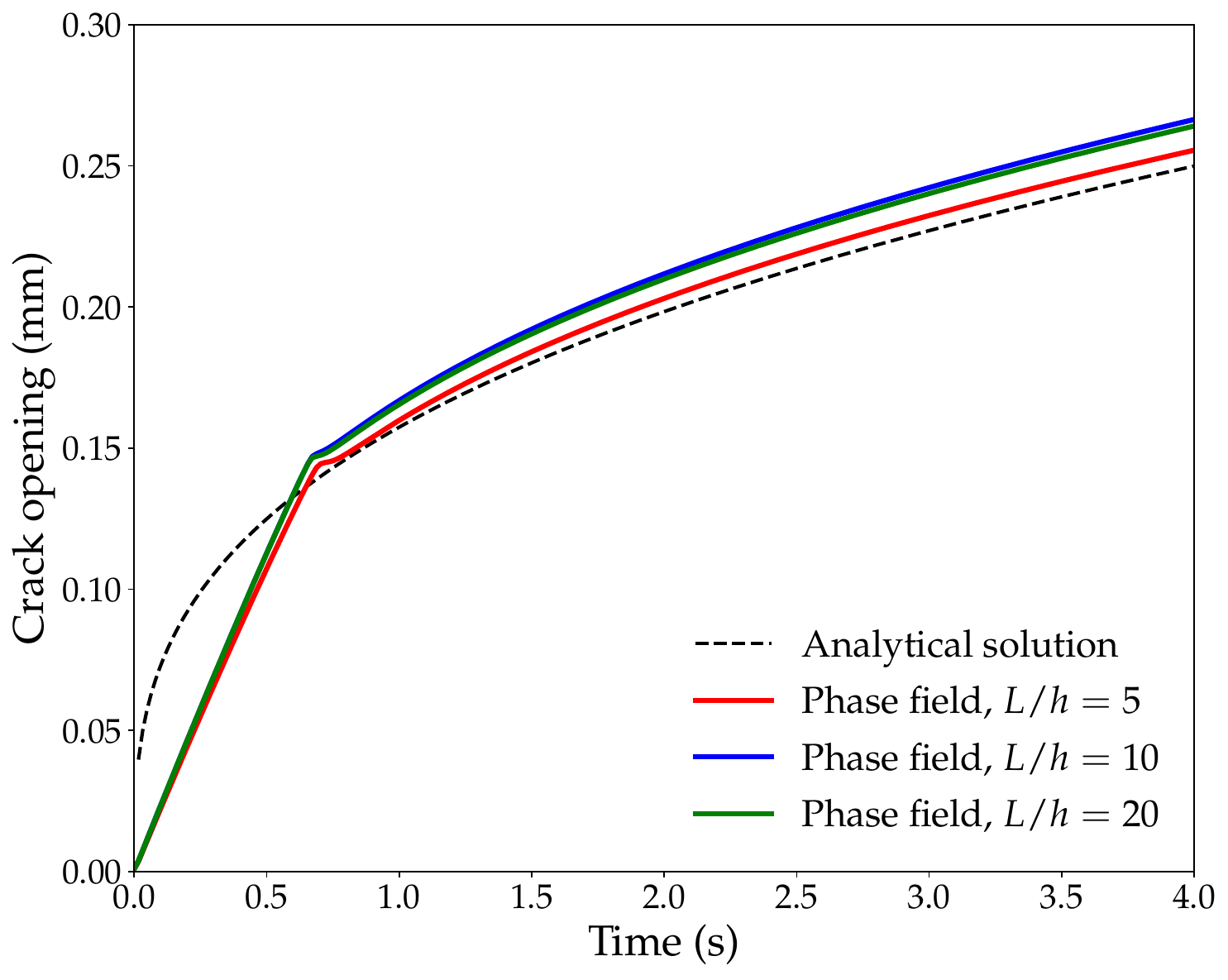} \label{fig:kgd-aperture-meshCompare}} \\
    \subfloat[]{\includegraphics[width=0.45\textwidth]{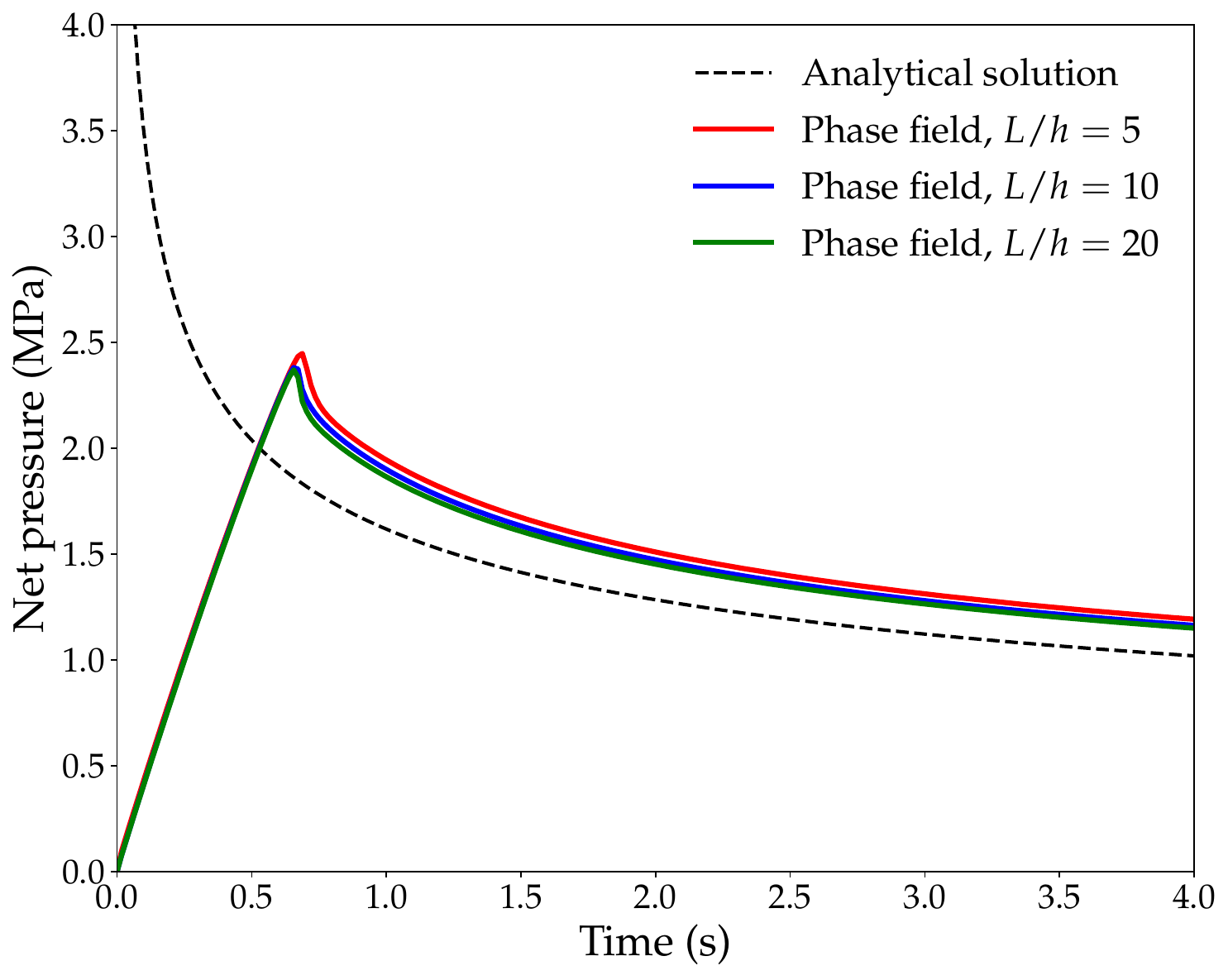} \label{fig:kgd-pressure-meshCompare}} \\ 
    \subfloat[]{\includegraphics[width=0.45\textwidth]{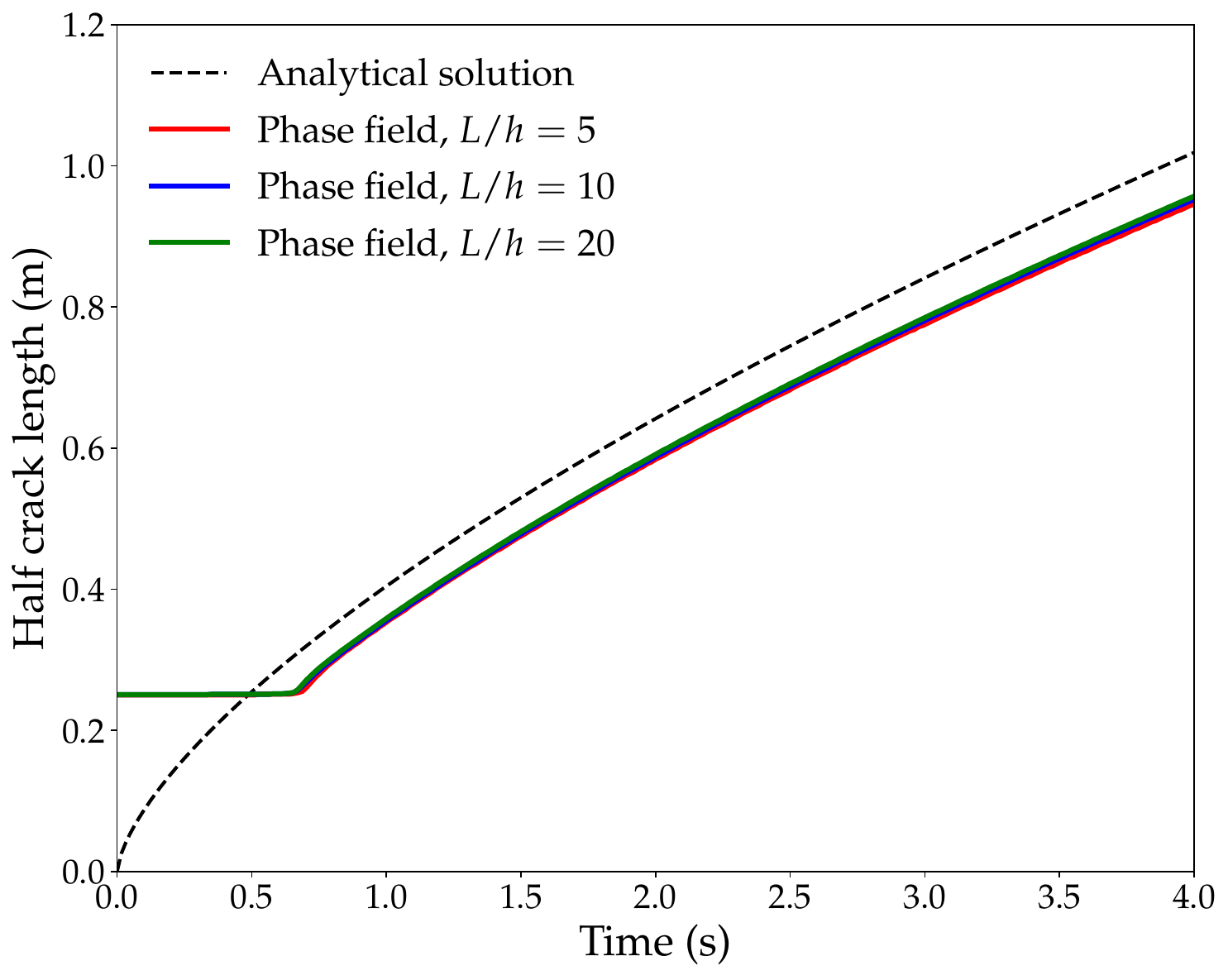} \label{fig:kgd-crackLength-meshCompare}}
    \caption{KGD hydraulic fracture propagation: comparison between the phase-field and analytical solutions in terms of the (a) crack opening at the center (b) net pressure at the crack center, and (c) half crack length, with different element sizes $h$. The regularization length is $L = 0.005$ m in all the cases.}
    \label{fig:kgd-meshCompare}
\end{figure}

In Fig.~\ref{fig:kgd-lengthCompare}, we present the simulation results in terms of the crack opening at the center, net pressure at the crack center, and half crack length, for different regularization lengths under a fixed discretization level of $L/h = 5$. 
Similarly, in Fig.~\ref{fig:kgd-meshCompare} we plot the phase-field solutions obtained with different levels of discretization while fixing the regularization length as $L = 0.005$ m. 
In these results, the crack opening and net pressure are measured at the material point closest to the crack center, while the half crack length is obtained by searching for the largest value of the $x$ coordinates of nodes where $d > 0.9$. 
One can see that the time evolution of the crack opening at the center---calculated by the proposed method---is in excellent agreement with the analytical solution. 
Further, the crack-opening solutions are consistent with changes in the phase-field regularization length and discretization level. 
Note that the proposed method works well even when the element size is extremely small, unlike the existing strain-based methods. 
While the phase-field solutions to the net pressure and half crack length show slight discrepancies with the analytical solutions, similar discrepancies have also been observed from other phase-field models for hydraulic fracture where the crack opening calculations are fully verified (\eg~\cite{chukwudozie2019variational,you2023poroelastic}).
So these discrepancies in the net pressure and half crack length are not directly related to the crack opening calculation which is the focus of the present work.
Therefore, it can be concluded that the proposed method can be used for calculating the crack openings of propagating phase-field fractures.

\section{Closure}
\label{sec:closure}

We have developed a strain-based method for calculating the crack opening of a phase-field fracture filled with fluid, which is critical to the phase-field modeling of fractures in coupled hydromechanical problems.
For the development, we have transformed the displacement-jump-based kinematics of a fracture into a continuous strain-based version, inserted it into a force balance equation on the fracture, and applied the phase-field approximation. 
Through this procedure, we have obtained a simple equation for the crack opening of a phase-field fracture, which can be calculated with quantities at individual material (quadrature) points.  
As such, the equation enables us to calculate the crack opening of a phase-field fracture without additional algorithms or parameters.
With benchmark examples in the literature, we have verified that the proposed method produces results practically identical to analytical and numerical solutions obtained based on discrete representations of fractures.
Also, the verification results confirm that the proposed method allows one to calculate the crack opening regardless of the element size or alignment, which is not the case for the existing strain-based calculations.
Considering that the difficulty of calculating the crack opening has been a major hurdle in the phase-field modeling of fluid-filled fractures, the proposed method is believed to significantly facilitate the applications of the phase-field method to modeling complex fractures in coupled hydromechanical problems.

\section*{Author Contributions} 
\label{sec:credit}

\textbf{Fan Fei}: Methodology, Software, Validation, Formal Analysis, Investigation, Writing - Original Draft, Visualization.
\textbf{Jinhyun Choo}: Conceptualization, Methodology, Validation, Writing - Original Draft, Writing - Review \& Editing, Supervision, Project Administration, Funding Acquisition.

\section*{Acknowledgements}
This work was supported by the National Research Foundation of Korea (NRF) grant funded by the Korean government (MSIT) (No. RS-2023-00209799).
Portions of this work were performed under the auspices of the U.S. Department of Energy by Lawrence Livermore National Laboratory under Contract DE-AC52-07NA27344.

\section*{Data Availability Statement} 
\label{sec:data-availability} 

The data that support the findings of this study are available from the corresponding author upon reasonable request.

\bibliography{references}

\end{document}